\pdfoutput=1

\documentclass[10pt,twocolumn,amsmath,amssymb,longbibliography,nofootinbib,showkeys,eprint]{revtex4-2}

\usepackage[activate={true,nocompatibility},final,tracking=true,kerning=true,spacing=true]{microtype}
\usepackage{epigraph}
\usepackage{graphicx}
\usepackage{epstopdf}
\usepackage{braket}
\usepackage{bm}
\usepackage{array}[=2016-10-06]
\usepackage{float}
\usepackage[T1]{fontenc}
\usepackage[utf8]{inputenc}
\usepackage{hhline}
\usepackage{tikz}
\usetikzlibrary{backgrounds,fit,decorations.pathreplacing,calc}
\usepackage[europeanresistors]{circuitikz}
\usepackage{xfrac}
\usepackage{amsthm}
\usepackage{makecell}
\usepackage{dsfont}
\usepackage{calc}
\usepackage{seqsplit}
\usepackage{hyperref}
\hypersetup{
    colorlinks=true
}
\usepackage{orcidlink}

\begin{document}
\preprint{V.M.}
\title{Superstate Quantum Mechanics}

\author{Mikhail Gennadievich \surname{Belov}}
\email{mikhail.belov@tafs.pro}
\affiliation{Lomonosov Moscow State University,  Faculty of Mechanics and Mathematics,
   GSP-1,  Moscow, Vorob'evy Gory, 119991, Russia}
\affiliation{Autretech Group, Skolkovo Innovation Center, Nobel Street, Building 7, Moscow, 121205, Russia}

\author{Victor Victorovich \surname{Dubov}}
\email{dubov@spbstu.ru}
\affiliation{Peter the Great St. Petersburg Polytechnic University, 195251, Russia}

\author{Vadim Konstantinovich \surname{Ivanov}\,\orcidlink{0000-0002-3584-4583}}
\email{ivvadim@rambler.ru}
\affiliation{Peter the Great St. Petersburg Polytechnic University, 195251, Russia}

\author{Alexander Yurievich \surname{Maslov}\,\orcidlink{0009-0005-6296-5988}}
\email{maslov.ton@mail.ioffe.ru}
\affiliation{Ioffe Institute, Politekhnicheskaya 26, St Petersburg, 194021, Russia}

\author{Olga Vladimirovna \surname{Proshina}\,\orcidlink{0000-0003-0087-5838}}
\email{proshina.ton@mail.ioffe.ru}
\affiliation{Ioffe Institute, Politekhnicheskaya 26, St Petersburg, 194021, Russia}

\author{Vladislav Gennadievich \surname{Malyshkin}\,\orcidlink{0000-0003-0429-3456}} 
\email{malyshki@ton.ioffe.ru}
\affiliation{Ioffe Institute, Politekhnicheskaya 26, St Petersburg, 194021, Russia}

\date{Dec, 19, 2024}

\begin{abstract}
\begin{verbatim}
$Id: SuperstateQuantumMechanics.tex,v 1.374 2026/06/25 18:21:12 mal Exp $
\end{verbatim}
We introduce Superstate Quantum Mechanics (SQM), a theory that considers
states in Hilbert space
subject to multiple quadratic constraints,
with ``energy'' also expressed as a quadratic function of these states.
Traditional quantum mechanics corresponds to a single quadratic constraint of wavefunction normalization
with energy expressed as a quadratic form involving the Hamiltonian.
When SQM represents states as unitary operators,
the stationary problem becomes a quantum inverse problem
with multiple applications in physics, machine learning, and artificial intelligence.
Any stationary SQM problem is equivalent to a new algebraic problem that we address in this paper.
The non-stationary SQM problem considers the evolution of the system itself,
involving the same ``energy'' operator as in the stationary case.
Two possible options for the SQM dynamic equation are considered:
(1) within the framework of linear maps from higher-order quantum theory; and
(2) in the form of a Gross-Pitaevskii-type nonlinear map.
This approach naturally bridges direct and inverse quantum mechanics problems,
allowing for the development of a new type of computer algorithms.
As an immediately available practical application of the theory,
we consider using a quantum channel as a classical computational model;
this type of computation can be performed on a classical computer.
\end{abstract}

\maketitle
\newpage

{\noindent Dedicated to the memory of Iya Pavlovna Ipatova, 1929--2003.}

\section{\label{intro}Introduction}

The structure of traditional quantum mechanics \cite{schrodinger1926quantisierung,born1926quantenmechanik}
involves defining a Hilbert space with the vectors $\Ket{\psi}$ and a Hamiltonian $H_{ij}$ represented as a Hermitian matrix.
The system’s energy in the state $\Ket{\psi}$ is then expressed as a quadratic form involving $H_{ij}$:
\begin{align}
\mathcal{F}&=
\sum\limits_{i,j=0}^{N-1}
\psi^*_i H_{ij} \psi_{j}
\label{FHamiltonian}
\end{align}
The Lagrange stationarity condition of (\ref{FHamiltonian}), subject to
the single quadratic constraint of wavefunction normalization
\begin{align}
1&=
\sum\limits_{i=0}^{N-1} |\psi_i|^2
\label{statePsi}
\end{align}
yields the stationary Schr\"{o}dinger equation in the form of an eigenvalue problem.
\begin{align}
\left|H\middle|\psi^{[i]}\right>&=\lambda^{[i]}\left|\psi^{[i]}\right> \label{stationarySchrodinger}
\end{align}
The time evolution of the wavefunction $\Ket{\psi}$ (a unit-length vector) in Hilbert space
is governed by the same operator $H_{ij}$, leading to the non-stationary Schr\"{o}dinger equation:
\begin{align}
i\hbar\frac{\partial \psi}{\partial t}&= H \psi
\label{SchrodingerOriginal}
\end{align}
A generalization of this dynamics defines a quantum channel \cite{nielsen2010quantum,wilde2011classical},
which is considered as a completely positive trace preserving (CPTP) map \cite{jamiolkowski1972linear,choi1975completely,kraus1983states,belavkin1986radon}.
The simplest example of such a channel is a unitary mapping.
\begin{align}
A^{OUT}&=\mathcal{U} A^{IN} \mathcal{U}^{\dagger}
\label{operatorTransformU}
\end{align}
For the Schr\"{o}dinger equation (\ref{SchrodingerOriginal}), the unitary operator $U$
\begin{align}
  U&=
  \exp \left[-i\frac{t}{\hbar} H \right] \label{Uquantum} \\
\Ket{\psi^{(t)}}&=\Ket{U \middle| \psi^{(0)}} \label{unitaryPsiEvolution}
\end{align}
defines a quantum channel (\ref{operatorTransformU}) that describes the time evolution of the initial state $\Ket{\psi^{(0)}}$,
this can be expressed by setting $\mathcal{U} = U$ and $A^{IN} = \Ket{\psi^{(0)}}\Bra{\psi^{(0)}}$.
The most commonly studied quantum channel is the one that describes the evolution of a quantum system between $t$ and $t+\tau$.
If the system has a time-independent Hamiltonian $H$, the evolution over a finite time can be obtained through multiple applications
of the quantum channel that describes the time evolution over a small interval $\tau$.
The time dependence of this quantum channel may arise solely from the explicit time dependence of the Hamiltonian, $H(t)$.

The key structural concepts of traditional quantum mechanics are
the definition of a vector space subject to the quadratic constraint (\ref{statePsi}),
the representation of energy as a quadratic form with $H$ (\ref{FHamiltonian}),
the derivation of an algebraic problem -- the eigenproblem (\ref{stationarySchrodinger}) -- as the stationary equation,
and the formulation of the non-stationary equation (\ref{SchrodingerOriginal}), which involves the same operator $H$.

There is an important generalization of the non-stationary equation,
from the evolution of a wavefunction to the evolution of the quantum system itself.
The most notable concept is that of higher-order quantum maps
\cite{chiribella2008transforming,bisio2019theoretical,odake2024higher,taranto2025higher},
which describe the evolution of an operator $U$ (\ref{Uquantum}) -- effectively, the ``dynamics of dynamics''.
The theory of higher-order quantum maps is primarily considered in the context of quantum computation
and is not commonly applied in classical computing.
These maps are typically studied as abstract transformations, without explicitly involving
a quadratic ``energy''-like functional that defines the mapping.

The main physical motivation of our work lies in the quantum inverse
problem\cite{bisio2010optimal,holzapfel2015scalable,bairey2019learning,li2020hamiltonian,stilck2024efficient,castaneda2025hamiltonian},
variational quantum algorithms \cite{cerezo2021variational,park2024hamiltonian,wang2024variationalPRL},
quantum tomography \cite{MAURODARIANO2003205,lvovsky2009continuous,torlai2023quantum,ahmed2023gradient},
and related areas.
From a set of observations (e.g., wavefunctions), one needs to reconstruct the underlying system
(e.g., determine its Hamiltonian).
In most cases, the problem reduces to finding a unitary operator that maximizes (or minimizes)
a functional (such as fidelity), subject to unitarity constraints.
Similar optimization problems arise in artificial intelligence
(see Ref. \cite{arjovsky2016unitary} and over 1000
\href{https://scholar.google.com/scholar?cites=5030720785335451277&as_sdt=2005&sciodt=0,5&hl=en}{subsequent works}),
where
knowledge representation also takes the form of a unitary operator,
leading to a constrained unitary optimization formulation.

These problems are essentially optimization problems
in a space of objects with properties richer than a unit-length vector (\ref{statePsi});
a unitary operator is a common choice.
In studies of quantum systems governed by the dynamics (\ref{unitaryPsiEvolution}),
the reconstructed unitary mapping can then be used to obtain the Hamiltonian
by taking the logarithm of the unitary matrix \cite{loring2014computing}.
There is no unique solution to Eq. (\ref{logUCalc});
many different Hamiltonians can be used to implement a quantum gate
$\mathcal{U}$ \cite{divincenzo1998quantum}.
\begin{align}
H&=i\frac{\hbar}{\tau} \ln \mathcal{U}
\label{logUCalc}
\end{align}
The specific problem of learning a unitary operator $\mathcal{U}$ from mapping data
is often referred to as ``tomography of unitary operations''
in quantum computation \cite{baldwin2014quantum, holzapfel2015scalable},
or ``unitary learning'' in artificial intelligence \cite{arjovsky2016unitary}.
We use the more general term ``quantum inverse problem'', which is also commonly employed
in the context of recovering Hamiltonians from scattering data as well as in quantum computations.
This choice reflects the fact that the constraints on the mapping operator can be more general than unitarity.
For example, partially unitary constraints, as considered in \cite{belov2024partiallyPRE}, are also valid within our framework.
Also, the entire field of machine learning can, in fact, be viewed as an inverse problem ---
reconstructing a model from data -- which is why we used the term ``inverse problem''.

There exists an alternative technique to learn the Hamiltonian as a sum of terms,
without first learning the unitary operator $\mathcal{U}$.
This approach is particularly advantageous for low-intersection Hamiltonians \cite{castaneda2025hamiltonian}.
However,
we will not pursue this route,
since the inverse problem in this form cannot be reduced to the algebraic problem we introduced.

The mentioned reconstruction techniques are optimization problems of some fidelity $\mathcal{F}$,
which is a general function of the mapping operators, for example a unitary operator $\mathcal{U}$.
The unitarity constraints, $\mathcal{U}^{\dagger}\mathcal{U}=\mathds{1}$,
constitute a set of $n(n+1)/2$ quadratic constraints on the mapping operators.
Note that the normalization condition, $1=\Braket{\psi|\psi}$ in (\ref{statePsi}), is a single quadratic constraint.

The main result of this work is the study of optimization problems in which the fidelity $\mathcal{F}$
is a quadratic function of the mapping operators.\footnote{
\label{notationsChange}
In this work, we changed the notation used in \cite{belov2024partiallyPRE,belov2024quantumPRE} from
$\Braket{a|b}=\sum_j a_j b^*_j$
to the common one in traditional quantum mechanics,
$\Braket{a|b}=\sum_j a^*_j b_j$.
}
\begin{align}
\mathcal{F}&=
\sum\limits_{j,j^{\prime}=0}^{D-1}\sum\limits_{k,k^{\prime}=0}^{n-1}
\mathcal{U}^*_{jk}
S_{jk;j^{\prime}k^{\prime}}
\mathcal{U}_{j^{\prime}k^{\prime}}
\label{Fidelity}
\end{align}
The mapping operators $\mathcal{U}_{jk}$ are $D\times n$ matrices subject to quadratic constraints.
Our work is not limited to unitary learning; it can be applied to any problem involving quadratic constraints,
such as partially unitary learning (\ref{unitarityCond}) for $D<n$,
potentially generalized to Kraus operator constraints (\ref{constraintKraussSpur}), and others.
While the quadratic form of $\mathcal{F}$ (\ref{Fidelity}) restricts the spectrum of problems considered,
it also allows a deeper understanding by enabling the formulation of
a new algebraic problem that generalizes the eigenvalue problem,
and the construction of a hierarchy of solutions.
Finally, we present an attempt to generalize the non-stationary  Schr\"{o}dinger equation.
The key similarity with traditional quantum mechanics is that both stationary and non-stationary
problems involve the same operator used to construct the quadratic form: $H_{jk}$ in (\ref{FHamiltonian})
and $S_{jk;j^{\prime}k^{\prime}}$ in (\ref{Fidelity}).
The superoperator $S_{jk;j^{\prime}k^{\prime}}$ serves as an analogue of the Hamiltonian $H_{ij}$.

The main computational motivation of this work lies in machine learning and artificial intelligence,
where problems with quadratic constraints naturally arise.
A limiting factor in the application of our theory lies in the requirement that the objective function (fidelity)
be quadratic in form for the algebraic approach to be applicable.
In \cite{belov2026semidefinitePRE}, we showed that the most general form of the fidelity
to which the theory can be applied is a ratio of two quadratic forms in the Kraus operators
${B}_{jk;s}$ (\ref{KrausOperator}):
\begin{align}
\mathcal{F}&=
\frac{
\sum\limits_{s=0}^{N_s-1}\sum\limits_{j,j^{\prime}=0}^{D-1}\sum\limits_{k,k^{\prime}=0}^{n-1}
{B}^*_{jk;s}
S_{jk;j^{\prime}k^{\prime}}
{B}_{j^{\prime}k^{\prime};s}
}
{
\sum\limits_{s=0}^{N_s-1}\sum\limits_{j,j^{\prime}=0}^{D-1}\sum\limits_{k,k^{\prime}=0}^{n-1}
{B}^*_{jk;s}
Q_{jk;j^{\prime}k^{\prime}}
{B}_{j^{\prime}k^{\prime};s}
}
 \xrightarrow[B]{\quad }\max
\label{fidelityProjectionsApproximationBBS}
\end{align}
For a unitary quantum channel (CPTP with Kraus rank one $N_s=1$), the numerator reduces to (\ref{Fidelity}),
while the denominator is equal to the sum (\ref{unitarityCondSimplified})
of all diagonal constraint elements in (\ref{constraintKraussSpur}).
This Rayleigh-quotient form of the fidelity allows the reconstruction of even non-TP mappings, including the important case
of reconstructing projective operators.
The modifications required when using the fidelity (\ref{fidelityProjectionsApproximationBBS})
are analogous to the transition from the standard eigenvalue problem
$\Ket{H|\psi}=\lambda \Ket{\psi}$ to the generalized eigenvalue problem
$\Ket{H|\psi}=\lambda \Ket{Q|\psi}$.
In this work, we consider the inverse problem from an algebraic perspective,
with the goal of obtaining a deeper understanding of the problem itself,
for which the use of a quadratic fidelity does not reduce the generality of the results.

When approaching the problem from a computational perspective,
there exists a fundamentally different approach to the optimization problem considered here:
Semidefinite Programming (SDP) \cite{vandenberghe1996semidefinite,wolkowicz2012handbook,bernard2009moments,anjos2011handbook,wen2013feasible,skrzypczyk2023semidefinite,tavakoli2024semidefinite,taranto2025higher}.
The key idea is to consider not the Kraus operators $B_s$ (\ref{KrausOperator}), but instead the Choi matrix:
\begin{align}
\mathcal{J}_{jk;j^{\prime}k^{\prime}}&=
\sum\limits_{s=0}^{N_s-1}
{B}^*_{jk;s}
{B}_{j^{\prime}k^{\prime};s}
\label{Jmatrix}
\end{align}
In this representation, maximizing (\ref{Fidelity}) subject to the quadratic TP constraints (\ref{constraintKraussSpur})
becomes the optimization problem (\ref{FidelityBSJJChoi})
subject to the linear constraints (\ref{constraintKraussSpurBSKKChoi}):
\begin{align}
\mathcal{F}&=\sum\limits_{j,j^{\prime}=0}^{D-1}\sum\limits_{k,k^{\prime}=0}^{n-1}
\mathcal{J}_{jk;j^{\prime}k^{\prime}}
S_{jk;j^{\prime}k^{\prime}} \xrightarrow[\mathcal{J}]{\quad }\max
\label{FidelityBSJJChoi} \\
\delta_{kk^{\prime}}&=\sum\limits_{j=0}^{D-1} \mathcal{J}_{jk;jk^{\prime}}
\label{constraintKraussSpurBSKKChoi}
\end{align}
All of these expressions are linear in the elements of the Choi matrix.
When the most general form of the fidelity (\ref{fidelityProjectionsApproximationBBS}) is used,
the resulting expressions remain linear and are modified slightly.

For a quantum channel to be completely positive (CP), an additional constraint is required:
the Choi matrix $\mathcal{J}$ must be positive semidefinite,
\begin{align}
\mathcal{J} &\succeq 0 \label{SDPconstraint}
\end{align}
Importantly, the optimization problem (\ref{FidelityBSJJChoi})
subject to the constraints (\ref{constraintKraussSpurBSKKChoi}) and (\ref{SDPconstraint}) is convex \cite{boyd2004convex}.
In \cite{belov2026semidefinitePRE}, we applied SDP to the numerical reconstruction of unitary operators,
quantum channels, projection operators, and other quantum mappings.
The convexity of the problem guarantees that the global optimum can be found
whenever the fidelity can be represented in the form (\ref{fidelityProjectionsApproximationBBS}).

In the case of unitary learning, SDP yields a rank-one Choi matrix $\mathcal{J}$ exactly,
from which the unitary operator $\mathcal{U}$ can be recovered as the eigenvector corresponding to its nonzero eigenvalue.
The convexity of the SDP formulation also explains why the iterative algorithm developed in \cite{belov2024partiallyPRE}
was able to find the global maximum for the unitary-learning problem.

In the present work, we focus not on the practical application of
SDP solvers \cite{yamashita2010high,andersen2015cvxopt,diamond2016cvxpy}
or other numerical techniques, but rather on the algebraic properties of higher-order dynamics.
Our goal is to construct, for a given higher-order dynamics, the corresponding algebraic problem --
a generalization of an eigenvalue problem -- and to understand its fundamental properties.
In Appendix \ref{QuadraticFidelity}, we have shown that the reconstruction of a general quantum channel
can also be reduced to a QCQP problem. However, applying this formulation requires computational optimization.

\section{\label{FidelityFormulation}Formulating Fidelity in Rayleigh-quotient Form}

The first step in approaching this problem is to introduce a stationary Schr\"{o}dinger-like equation
that describes a quantum inverse problem.
This stationary problem has a clear physical meaning: recovering the system Hamiltonian from a measured sequence of system states,
and it arises in multiple physical and machine learning applications \cite{bisio2010optimal,arjovsky2016unitary}.
Consider the quantum inverse problem
as an optimization problem.
For a given sequence of observations $l=1\dots M$, of pure (\ref{purePsiMap}) or mixed states (\ref{mixedRhoMap})
mapping,
\begin{align}
\psi^{(l)} &\to \phi^{(l)} \label{purePsiMap} \\
\rho^{(l)}  &\to \varrho^{(l)} \label{mixedRhoMap}
\end{align}
 the goal is to reconstruct the unitary operator $\mathcal{U}$ (\ref{operatorTransformU}) that optimally maps
$A^{IN}=\Ket{\psi^{(l)}}\Bra{\psi^{(l)}}$
to $A^{OUT}=\Ket{\phi^{(l)}}\Bra{\phi^{(l)}}$, or for mixed states,
 $A^{IN}=\rho^{(l)}$ to $A^{OUT}=\varrho^{(l)}$ 
for all $l=1\dots M$.
By interpreting $l$ as time $t$ (where $l\to l+1$ corresponds to $t\to t+\tau$),
the time evolution of a quantum system, $\psi^{(l)} \to \psi^{(l+1)}$,
or, for mixed states, $\rho^{(l)} \to \rho^{(l+1)}$,
can be described by setting $\phi^{(l)}=\psi^{(l+1)}$ or $\varrho^{(l)}=\rho^{(l+1)}$, respectively.

Our algebraic approach is commonly applicable when the objective function $\mathcal{F}$\footnote{
\label{footnoteFidelity}
The objective function is typically the total fidelity \cite{nielsen2010quantum,wilde2011classical,wilde2018recoverability} of the mapping, for other approaches, see \cite{budini2024quantum}.
The main requirement for ``alternative fidelity'' is that it reaches its maximum when the mapping is equal to the exact quantum channel;
this requirement is analogous to Uhlmann's theorem for quantum-channel fidelity, but is applied to maps that are completely positive (CP) and not necessarily trace-preserving (TP).
See Table I of Ref. \cite{belov2024quantumPRE} for counterexamples where,
for some fidelity proxies (approximations),
there exists a single unitary mapping that yields a proxy-fidelity
greater than that of the full general quantum channel.
Variational quantum algorithms \cite{cerezo2021variational,park2024hamiltonian,wang2024variationalPRL} similarly use a cost function that is quadratic in $\mathcal{U}$, typically of the form
$\mathcal{F}= \mathrm{Tr}\, O \mathcal{U}\rho_0\mathcal{U}^{\dagger}$,
which yields the tensor
$S_{jk;j^{\prime}k^{\prime}}=O_{jj^{\prime}}\left(\rho_0\right)_{k^{\prime}k}$.
}
can be represented as a ratio of two quadratic forms (\ref{fidelityProjectionsApproximationBBS})
in the operator $\mathcal{U}$, typically unitary or partically unitary.

In a reconstruction problem involving a general mixed-state mapping (\ref{mixedRhoMap}),
the total fidelity is defined as the sum of the weighted mapping accuracies over all $M$ observations
(throughout this section (and in what follows), we assume for simplicity that all weights $\omega^{(l)}=1$;
different weights are often used in AI applications to reduce the contribution of high-error measurements).
\begin{align}
\mathcal{F}&=\sum_{l=1}^{M} \omega^{(l)}
F(\varrho^{(l)},\sigma^{(l)})
\label{fidelityStdDefinitionGeneralCase}
\end{align}
Here, $F(\varrho,\sigma)$ is a measure of closeness between the states $\varrho$ and $\sigma$, such as fidelity.
The matrix $\sigma^{(l)}$ denotes the density matrix $\rho^{(l)}$ after it has been passed through a quantum channel,
whether it is given by (\ref{operatorTransformU}) or (\ref{KrausOperator}).
With the fidelity between mixed states $\varrho$ and $\sigma$, see
\cite{nielsen2010quantum} p. 409,
\cite{wilde2011classical} p. 285,
obtain
\begin{align}
F(\varrho,\sigma)&=\left|\mathrm{Tr}\sqrt{\varrho^{1/2}\sigma\varrho^{1/2}}\right|^2=
\left(\mathrm{Tr}\left|\sqrt{\varrho}\sqrt{\sigma}\right|\right)^2
\label{fidelityDefinitionTextBook}
\end{align}
From a computational perspective, it is beneficial to use, insead of (\ref{fidelityDefinitionTextBook}),
the Holevo fidelity, Eq. (9.249) in Ref. \cite{wilde2011classical}:
\begin{align}
F_H(\varrho,\sigma)&=
\left(\mathrm{Tr}\sqrt{\varrho}\sqrt{\sigma}\right)^2
\label{fidelitySQRTvarrosigma}
\end{align}
Although this form simplifies certain computational aspects,
it is generally not equivalent to (\ref{fidelityDefinitionTextBook}), except in the case of a unitary mapping.

In the general case, the total fidelity $\mathcal{F}$ (\ref{fidelityStdDefinitionGeneralCase})
is not necessarily quadratic in the quantum channel mapping operators
and cannot be represented as (\ref{Fidelity}).
However, the problem becomes particularly simple for unitary mapping of pure states (\ref{purePsiMap}), where
\begin{align}
\mathcal{F}&=\sum_{l=1}^{M} \omega^{(l)}
\left|\Braket{\phi^{(l)}|\mathcal{U}|\psi^{(l)}}\right|^2
\label{fidelityDefinitionPure}
\end{align}
is a quadratic function of $\mathcal{U}$.
Expanding the parentheses yields a simple expression for $S_{jk;j^{\prime}k^{\prime}}$.
\begin{align}
S_{jk;j^{\prime}k^{\prime}}&=
\sum\limits_{l=1}^{M}\omega^{(l)} \phi^{(l)*}_j\phi^{(l)}_{j^{\prime}}\psi^{(l)*}_k\psi^{(l)}_{k^{\prime}}
\label{SexamplePureExact}
\end{align}
For mixed states (\ref{mixedRhoMap}) general mapping (\ref{KrausOperator}),
the situation is more complicated, and no general method is currently available
for expressing fidelity as a quadratic function of the mapping operators.
Various approximations (proxies) can be used to obtain a quadratic form of fidelity,
but this approach has its own limitations \cite{belov2024quantumPRE}.
The situation is somewhat simpler for mixed-state unitary mappings (\ref{operatorTransformU}),
where the same quantum channel transforms both $\rho$ and $\sqrt{\rho}$.
One may consider
\begin{align}
S_{jk;j^{\prime}k^{\prime}}&=
\sum\limits_{l=1}^{M}\omega^{(l)} \widetilde{\varrho}^{(l)*}_{jj^{\prime}}\widetilde{\rho}^{(l)*}_{kk^{\prime}}
\label{SexampleMixedExact}
\end{align}
where $\widetilde{\varrho}$ and $\widetilde{\rho}$ are simple functions of $\varrho$ and $\rho$.
This corresponds to considering
$\widetilde{\rho}=\sqrt{\rho}\to\sqrt{\varrho}=\widetilde{\varrho}$
instead of the original $\widetilde{\rho}=\rho\to\varrho=\widetilde{\varrho}$.
The issue with this approach is that it corresponds to the fidelity
$F(\varrho,\sigma)=\mathrm{Tr}\sqrt{\varrho}\sqrt{\sigma}$,
as in Eq. (\ref{fidelitySQRTvarrosigma}), without taking the absolute value $|\cdot|$
and using the ``square root fidelity''.
In general, $\mathrm{Tr}\left(\sqrt{\varrho}\sqrt{\sigma}\right)\ne\mathrm{Tr}\left|\sqrt{\varrho}\sqrt{\sigma}\right|$.
As a result, in the case of unitary mappings of mixed states, the expression (\ref{SexampleMixedExact}) underestimates the true fidelity.
However, it remains quadratic in  $\mathcal{U}$  and can be considered as an approximation \cite{belov2024quantumPRE}.
Numerical experiments show that it differs from the true fidelity only when the fidelity value is low;
for high values, the two expressions match almost exactly.
It is important to note that for unitary mapping of mixed states
this fidelity approximation satisfies the property mentioned in footnote \ref{footnoteFidelity}
---  it reaches its maximum when the mapping is equal to the exact.

The form of $S$ given by (\ref{SexampleMixedExact}) (or (\ref{SexamplePureExact}) for pure states)
has an interesting relationship to an autocorrelation function
of a dynamic system.
For a dynamic system where $\varrho^{(l)}=\rho^{(l+1)}$,
the calculation of $S_{jk;j^{\prime}k^{\prime}}$ is effectively reduced to an autocorrelation
 with a delay of $\tau$
\begin{align}
S_{jk;j^{\prime}k^{\prime}}&=
\sum\limits_{l=1}^{M}\omega^{(l)} \widetilde{\rho}^{(l+1)*}_{jj^{\prime}}\widetilde{\rho}^{(l)*}_{kk^{\prime}}
\label{SexampleMixedExactCorrelation}
\end{align}
If it were a classical system,
this calculation would be equivalent to measuring or computing a time average.
In quantum systems, however, the situation is more complex since the measurement of $\rho^{(l)}$ destroys the state,
making $\rho^{(l+1)}$ at the next step potentially unavailable.
A question arises as to whether the entire ``time average'' (\ref{SexampleMixedExactCorrelation})
can be treated as the result of a single act of quantum measurement,
without requiring observation of the system state $M$ times at each specified time $t=l\tau$.
Recently, new approaches have been developed that allow for the measurement of even multiple-time autocorrelations
in quantum systems\cite{wang2024snapshotting};
in this approach, $S_{jk;j^{\prime}k^{\prime}}$ is obtained from the single-time density matrix autocorrelation
through continuous observation of the system state.

Alternatively, instead of continuous observation of the system state, one can obtain the  $S_{jk;j^{\prime}k^{\prime}}$
using a process typical for quantum computations.
In quantum computations, an initial state $\Ket{\psi^{(0)}}$ is prepared in a specific state of qubits.
Then, a unitary transformation corresponding to the required quantum circuit is applied, and the result is measured.
Similarly, we can create $\Ket{\psi^{(0)}}$ randomly and measure the result $\Ket{\psi^{(\tau)}}$ of the system's evolution.
In this approach, instead of continuously observing the system,
we randomly create the initial state $M$ times and measure the corresponding result of its evolution after a time duration of $\tau$.
This process may be easier to implement than continuous observation of a quantum system's state.
It is also simpler than the protocol of quantum computation,
as $\Ket{\psi^{(0)}}$ is taken randomly and does not need to be prepared in a specific state.
The only requirement for the set of $M$ mappings is that it be information-complete\cite{torlai2023quantum}.

Up to this point, the total fidelity has been considered as a single sum (\ref{fidelityStdDefinitionGeneralCase})
of fidelity contributions corresponding to individual observations.
In the case of perfect matching, this yields a fidelity equal to the number of observations in (\ref{purePsiMap}), namely
$\mathcal{F}=M$, when all weights $\omega^{(l)}=1$.
Whenever the resulting sum is quadratic in the mapping operator  $\mathcal{U}$ (probe operator),
the algebraic analysis developed in this work can be applied directly. Otherwise, some form of fidelity approximation is required.
The problem of selecting an appropriate fidelity measure (objective function)
is a broad area of research \cite{wilde2018recoverability,budini2024quantum,johnston2011quantum}.
We do not attempt to discuss all possible choices here.
Rather, we wish to demonstrate that our theory applies to a broad class of fidelity functions of the Rayleigh-quotient type.
Specifically, instead of considering a single quadratic form,
one may construct two quadratic functions of the mapping operator and consider their ratio.
This more general form can also be treated within the theoretical framework developed here.
The numerator and denominator are typically obtained as sums of terms that are quadratic in the mapping operator,
although the individual terms do not themselves have the interpretation of a fidelity.
This is the approach pursued in \cite{belov2026semidefinitePRE}
to derive the most general form (\ref{fidelityProjectionsApproximationBBS}).
For example, in the reconstruction of a projection operator (Kraus rank $N_s=1$)
from pure-state-to-pure-state mapping data (\ref{purePsiMap}), the fidelity takes the form
\begin{align}
\mathcal{F}&=
\frac{
\sum\limits_{j,j^{\prime}=0}^{D-1}\sum\limits_{k,k^{\prime}=0}^{n-1}
\mathcal{U}^*_{jk}
S_{jk;j^{\prime}k^{\prime}}
\mathcal{U}_{j^{\prime}k^{\prime}}
}
{
\sum\limits_{j,j^{\prime}=0}^{D-1}\sum\limits_{k,k^{\prime}=0}^{n-1}
\mathcal{U}^*_{jk}
Q_{jk;j^{\prime}k^{\prime}}
\mathcal{U}_{j^{\prime}k^{\prime}}
}
 \xrightarrow[\mathcal{U}]{\quad }\max
\label{fidelityProjectionsApproximationBBSProjectionsU}
\end{align}
where $\mathcal{U}$ is the $D\times n$ projection operator to be reconstructed.
The tensor $S_{jk;j^{\prime}k^{\prime}}$ is given by (\ref{SexamplePureExact}), while $Q_{jk;j^{\prime}k^{\prime}}$ is
\begin{align}
Q_{jk;j^{\prime}k^{\prime}}&=
\delta_{jj^{\prime}}\sum\limits_{l=1}^{M}\omega^{(l)} \psi^{(l)*}_{k} \psi^{(l)}_{k^{\prime}}
\label{QexamplePurePureExact}
\end{align}
The algebraic approach developed in this work can be applied to fidelities
of the form (\ref{fidelityProjectionsApproximationBBSProjectionsU})
for arbitrary tensors $S_{jk;j^{\prime}k^{\prime}}$ and $Q_{jk;j^{\prime}k^{\prime}}$.

For objective functions of the Rayleigh-quotient type and quadratic constraints (such as CPTP, partially unitary, and related constraints), the problem becomes a Quadratically Constrained Quadratic Program (\href{https://en.wikipedia.org/wiki/Quadratically_constrained_quadratic_program}{QCQP}).
Only in this case is the algebraic approach developed in this work applicable,
allowing the derivation of a stationary Schr\"{o}dinger-like equation (\ref{eigenvaluesLikeProblem}).
In the non-quadratic case, the problem becomes a general optimization task lacking a specific algebraic structure.
It can then only be addressed numerically using gradient-based optimization methods,
which generally do not guarantee finding the global maximum or determining the complete hierarchy of solutions.
The simplest example of a stationary problem in traditional quantum mechanics is the determination of the ground state,
namely the unit-norm vector $\Ket{\psi}$ that minimizes the energy.
The energy is expressed as the quadratic form (\ref{FHamiltonian}) in $\Ket{\psi}$ with Hamiltonian $H$,
and the corresponding optimization problem reduces to the eigenvalue problem (\ref{stationarySchrodinger}).
In contrast, the simplest example of a stationary problem in SQM is the determination of a  unitary operator $\mathcal{U}$
(or, more interestingly, a projection operator when $D<n$) that maximizes the fidelity subject to the corresponding (partial) unitarity constraints.

This paper is accompanied by a software which
\href{http://www.ioffe.ru/LNEPS/malyshkin/code_polynomials_quadratures.zip}{is available}
from Ref. \cite{polynomialcode};
all references to code in the paper correspond to this software.

\section{\label{stationary}A stationary Schr\"{o}dinger-like equation}

In traditional quantum mechanics, a quantum state (pure state case) is a unit-length vector $\Ket{\psi}$ in
\href{https://en.wikipedia.org/wiki/Hilbert_space}{Hilbert space}.
Quantum dynamics describes the time evolution of this vector within it.
The Hilbert space is formally defined as an infinite-dimensional vector space (with a countable number of basis vectors)
equipped with an inner product.
However, we also refer to finite-dimensional spaces as Hilbert spaces in this context.
The inner product $\Braket{\psi|\phi}$ can be defined as
\begin{align}
\Braket{\psi|\phi}&=
\sum\limits_{j,k=0}^{N-1} \psi_j^* Q_{jk} \phi_k
\label{scalProdQ}
\end{align}
with an arbitrary positively definite Hermitian matrix $Q_{jk}$. It can be reduced to a familiar form
\begin{align}
\Braket{\psi|\phi}&=
\sum\limits_{i=0}^{N-1} \psi_i^* \phi_i
\label{scalProd}
\end{align}
through a basis linear transformation.
Traditional quantum mechanics consists of finding a stationary state $\Ket{\psi}$ or its time evolution $\Ket{\psi^{(t)}}$,
where the state is subject to a single quadratic constraint of wavefunction normalization (\ref{statePsi}).

In our previous works \cite{belov2024partiallyPRE,belov2024quantumPRE} on the quantum inverse problem, states with multiple quadratic constraints naturally arise.
The simplest form of this type of state is a unitary (or partially unitary) operator $\mathcal{U}$,
represented by the matrix $\mathcal{U}_{jk}$, where $j = 0 \dots D-1$, $k = 0 \dots n-1$, and $D \leq n$, satisfying the constraints.
\begin{align}
G_{ij}     &=\sum\limits_{k=0}^{n-1} \mathcal{U}_{ik} \mathcal{U}^*_{jk} & i,j=0\dots D-1
\label{GramMatrix} \\
\delta_{ij}&=G_{ij}
\label{unitarityCond}
\end{align}
The Gram matrix (\ref{GramMatrix}),
introduced as the partial convolution of $\mathcal{U}_{jk}$, is a unit matrix for partially unitary $\mathcal{U}_{jk}$.
When $D = n$, condition (\ref{unitarityCond}) represents the unitarity of $\mathcal{U}_{jk}$.
In this case, it is equivalently expressed as Eq. (\ref{constraintKraussSpur}) for $N_s=1$,
\begin{align}
\delta_{kq}&=\sum\limits_{j=0}^{D-1} \mathcal{U}^*_{jk} \mathcal{U}_{jq} & k,q=0\dots n-1
\label{equivUnitaryCond}
\end{align}
but this form with $N_s=1$ cannot be satisfied when $D<n$.
However, condition (\ref{unitarityCond}) allows this case to be studied.
The unitary operator $\mathcal{U}$ creates a transformation of an arbitrary operator $A$,
acting as a quantum channel (\ref{operatorTransformU}).

The constraints in (\ref{unitarityCond}) can be conveniently interpreted
as the transformation of the unit matrix $A^{IN}$ into the unit matrix $A^{OUT}$
using the quantum channel (\ref{operatorTransformU});
this type of constraint is especially important when reconstructing projection operators.
When $D=n$, the quantum channel represents a trace-preserving map, which can describe, for example, the evolution of a density matrix $\rho_{jk}$ for a quantum system.
When $D<n$, this is a trace-decreasing map,
which is typically not used in the study of quantum systems but may be very beneficial
for exploring numerical algorithms that reconstruct projection operators.
When considering a general quantum channel\cite{kraus1983states,gyongyosi2012properties}
\begin{align}
A^{OUT}&=\sum_{s=0}^{N_s-1} B_s A^{IN} B^{\dagger}_s
\label{KrausOperator}
\end{align}
the constraints on $B_s$ corresponding to the trace preservation of the mapping take the form:
\begin{align}
  \sum\limits_{s=0}^{N_s-1} B_s^{\dagger}B_s&=\mathds{1} \label{constraintKraussSpur}
\end{align}
However, in this work, we primarily focus on (partially) unitary channels (\ref{operatorTransformU})
with Kraus rank $N_s=1$.
Partial unitarity constraints (\ref{unitarityCond}) comprise $D$ conditions for the diagonal elements to be equal to one and
$D(D-1)/2$ conditions for the off-diagonal elements to be equal to zero.
The inverse problem consists of identifying the operator $\mathcal{U}$
that maximizes the fidelity subject to a set of quadratic constraints on $\mathcal{U}$.

\subsection{\label{NumSolutionExisting}Computational Approaches to the Optimization Problem}
There are two distinct approaches to this optimization problem.
One is based on semidefinite programming (SDP) and is discussed in \cite{belov2026semidefinitePRE}.
The other is an iterative approach based on the solution of eigenvalue problems \cite{belov2024partiallyPRE}.
We summarize its basic properties here, as they help elucidate the underlying algebraic structure of the problem.

First, the full set of $D(D+1)/2$ constraints (\ref{unitarityCond}) is transformed into $D(D+1)/2-1$
homogeneous constraints --- $D(D-1)/2$ requiring the off-diagonal elements to be zero and $D-1$
requiring all diagonal elements to be equal ---
and a single inhomogeneous constraint requiring the sum of the diagonal elements to be constant.
This latter constraint is referred to as the ``simplified'' (partial)
quadratic constraint and is of particular importance for the numerical method.\footnote{
The transformation $T$, which converts an operator $\mathcal{U}$ satisfying Eq. (\ref{unitarityCondSimplified})
into a unitary operator $\widetilde{\mathcal{U}}=T\mathcal{U}$ satisfying Eq. (\ref{unitarityCond}),
is used in the numerical method discussed in ``Appendix A2. Adjustment of operators to orthogonal''
of Ref. \cite{belov2024quantumPRE}.
This transformation corresponds to $TGT^{\dagger}=\mathds{1}$.
Since $G=\mathcal{U}\mathcal{U}^{\dagger}$ is Hermitian, the problem reduces to calculating the inverse square root of $G$,
 $T=T^{\dagger}=G^{-1/2}$.
}
\begin{align}
D&=\sum\limits_{j=0}^{D-1}\sum\limits_{k=0}^{n-1} |\mathcal{U}_{jk}|^2
\label{unitarityCondSimplified}
\end{align}
When the Rayleigh-quotient fidelity (\ref{fidelityProjectionsApproximationBBSProjectionsU})
is employed in place of (\ref{Fidelity}),
the partial constraint becomes the requirement that the denominator of (\ref{fidelityProjectionsApproximationBBSProjectionsU})
be equal to a constant.
\begin{align}
\mathrm{const}&=
\sum\limits_{j,j^{\prime}=0}^{D-1}\sum\limits_{k,k^{\prime}=0}^{n-1}
\mathcal{U}^*_{jk}
Q_{jk;j^{\prime}k^{\prime}}
\mathcal{U}_{j^{\prime}k^{\prime}}
\label{unitarityCondSimplifiedQ}
\end{align}
If we were to consider only the partial constraint,
this would effectively be a regular eigenproblem (\ref{statePsi}), with the vector $\psi_i$
obtained from the operator $\mathcal{U}_{jk}$ by
\href{https://en.wikipedia.org/wiki/Vectorization_(mathematics)}{vectorization},
saving all $\mathcal{U}_{ik}$  matrix elements into a single vector, row by row,
to obtain a vector $\psi_i$ of dimension $N = Dn$.
Note that when only the partial constraint is imposed,
the largest eigenvalue of the resulting eigenproblem provides an estimate
of the maximal achievable fidelity for a general quantum channel.
Since additional constraints can only reduce the feasible set,
they can only decrease the obtained value.
Therefore, the eigenproblem solution provides an upper bound on the fidelity in the inverse problem.

Considering vectors in Hilbert space subject to multiple quadratic constraints (e.g., a unitary operator $\mathcal{U}$)
naturally bridges direct and inverse quantum-mechanical problems.
In this framework, quantum states are generalized to operators,
and the operators acting on them become superoperators.
With only a single quadratic constraint -- (\ref{statePsi}), i.e., unit-norm vectors --
we recover standard quantum mechanics as a limiting case.
SQM can be viewed as an approach that considers system states
at the intersection of manifolds,
each being a second-order surface.
The number of surfaces that intersect equals the number of constraints in (\ref{unitarityCond}),
and their intersection is not always transversal.
The iterative method developed for finding the global maximum \cite{belov2024partiallyPRE,belov2024quantumPRE}
involves replacing, at each iteration,
the $D(D+1)/2$ quadratic constraints (\ref{unitarityCond})
with a single simplified quadratic constraint
and $D(D+1)/2-1$ homogeneous linear constraints.\footnote{
If there were only a single solution candidate, the chances of finding the global maximum, with some constraints linearized,
would be comparable to using Newton's method, which, at each iteration,
has a single solution candidate obtained from solving a linear system.
In contrast, using the eigenproblem at each iteration, which has multiple solution candidates (eigenvectors), alters the situation.
}
Based on numerical experiments, we conclude that the iterative algorithm almost always converges to the global maximum.
Whether this behavior can be formally established as a theorem remains an open question for future research.
The current implementation is available for real-space matrices with arbitrary quadratic constraints (not limited to unitary).
The algorithm for complex numbers has not been developed yet,
since the main application areas of this theory -- machine learning and artificial intelligence -- typically do not require it.
The optimization algorithm was extensively tested on a large amount of data,
including both actual machine learning datasets and randomly generated unitary and partially unitary data.
The algorithm demonstrates almost perfect convergence for the first solution in the hierarchy;
subsequent solutions may not always be found due to the issues described in Appendix \ref{SolutionsHierarchyConstraints}.
All code and scripts are available from \cite{polynomialcode}.
This unitary optimization is a form of QCQP problem.
There are several QCQP software packages available, both commercial and open-source, such as 
\cite{frison2022introducing},
\href{https://github.com/cvxgrp/qcqp}{python qcqp},
\href{https://nag.com/solving-quadratically-constrained-quadratic-programming-qcqp-problems/}{NAG QCQP}
among many
\href{https://en.wikipedia.org/wiki/Quadratically_constrained_quadratic_program\#Solvers_and_scripting_(programming)_languages}{others}.
The major limitation of these packages is that they are designed to solve \textsl{arbitrary} QCQP problems and often handle the specific problem considered here poorly.
The main reason for their poor performance is that these methods are based on gradient- or Newton-type algorithms,
which generate only a single solution candidate.
In contrast, our approach uses an eigenvalue problem as the core building block,
naturally producing multiple solution candidates (eigenvectors).
Combined with explicitly written homogeneous linear constraints, this allows for more efficient and accurate computation.
These linear constraints, referred to as convergence-helper constraints,
are obtained as tangents to the constraint surfaces at the point of the current iteration.
Traditional quantum mechanics corresponds to considering states on the unit sphere,
greatly simplifying the matter.
Similarly, it is more convenient to develop the theory in a regular
Hilbert space with multiple quadratic constraints (\ref{unitarityCond}),
rather than in the space of intersections of multiple second-order surfaces.

\subsection{\label{algebraicProblemOpt}Algebraic Representation of the Optimization Problem}
After defining the system state with the unitary operator $\mathcal{U}_{jk}$,
we introduce a quadratic form used in a Hamiltonian-like manner.
In traditional quantum mechanics, with the single constraint (\ref{statePsi}),
this would correspond to a familiar quadratic expression (\ref{FHamiltonian}), where the Hamiltonian $H_{ij}$ is represented as a Hermitian matrix.
When a quantum ``superstate'' is defined by a unitary matrix $\mathcal{U}_{jk}$ (e.g., in the quantum inverse problem),
the total fidelity of the pure-state mapping (\ref{purePsiMap}),
$\psi^{(l)} \to \phi^{(l)}$, $l=1\dots M$,
can be expressed in quadratic form (\ref{Fidelity}) with an appropriate tensor $S_{jk;j^{\prime}k^{\prime}}$.
For partially unitary superstate $\mathcal{U}_{jk}$ and the Rayleigh-quotient fidelity (\ref{fidelityProjectionsApproximationBBSProjectionsU}),
two approaches can be considered:
1. The tensor $Q_{jk;j^{\prime}k^{\prime}}$ from (\ref{QexamplePurePureExact}) can be reduced to the form $\delta_{jj^{\prime}}\delta_{kk^{\prime}}$
by transforming the input vectors  $ \psi^{(l)}$ (\ref{purePsiMap}) to a basis in which the matrix
$\sum\limits_{l=1}^{M}\omega^{(l)} \psi^{(l)*}_{k} \psi^{(l)}_{k^{\prime}}$
is diagonal.
In this case, the denominator in (\ref{fidelityProjectionsApproximationBBSProjectionsU}) can be omitted,
since it becomes absorbed
into the sum of the diagonal elements of the partially unitary constraints (\ref{unitarityCond}).
2. If $Q_{jk;j^{\prime}k^{\prime}}$ has a more general form than (\ref{QexamplePurePureExact}),
one may instead use the simplified constraint (\ref{unitarityCondSimplifiedQ}) together with the $D(D+1)/2-1$ homogeneous constraints,
as discussed above for the numerical method. This again leads to a similar linear algebraic problem,
albeit with more complex expressions.
In this section, we restrict ourselves to (\ref{QexamplePurePureExact}), for which one can assume, after an appropriate basis transformation, $Q_{jk;j^{\prime}k^{\prime}}=\delta_{jj^{\prime}}\delta_{kk^{\prime}}$,
which makes it sufficient to consider the quadratic fidelity form (\ref{Fidelity}).

The ground state stationary solution of the system with $S_{jk;j^{\prime}k^{\prime}}$
consists in finding the global maximum (or minimum for some other problems) of (\ref{Fidelity})
with $\mathcal{U}_{jk}$ subject to (\ref{unitarityCond}) constraints.
For an evaluation of the computational complexity of our algebraic-type algorithm,
see \cite{belov2024quantumPRE}.\footnote{
For a general Hermitian matrix eigenproblem of dimension $N$, the computational complexity is $O(N^3)$ \cite{demmel2007fast}.
Similarly, when $D=n$ and hence $N=n^2$,
the problem of finding all solutions to equation (\ref{eigenvaluesLikeProblemSlev})
has a computational complexity of $O(n^6)$ for determining all matrix pairs $\lambda^{[s]},\mathcal{U}^{[s]}$.
However, in most cases, only a single ``ground state'' solution is needed.
Using certain techniques \cite{demmel2007fast,belov2024partiallyPRE},
the computational complexity of this problem can likely be reduced to $O(n^4)$.
This is roughly on par with gradient-based methods,
which are typically faster but do not guarantee finding the global maximum.
}

Let us reformulate the QCQP optimization problem of maximizing (\ref{Fidelity}) into an algebraic form.
By constructing the Lagrangian $\mathcal{L}$ with the constraints (\ref{unitarityCond}) conjugated,
and varying it with respect to $\mathcal{U}_{jk}$,
\begin{align}
  \mathcal{L}&=
  \sum\limits_{j,j^{\prime}=0}^{D-1}\sum\limits_{k,k^{\prime}=0}^{n-1}
             \mathcal{U}^*_{jk}S_{jk;j^{\prime}k^{\prime}}\mathcal{U}_{j^{\prime}k^{\prime}} \nonumber \\
             &+
   \sum\limits_{j,j^{\prime}=0}^{D-1}        
   \lambda_{jj^{\prime}}\left[\delta_{jj^{\prime}}-\sum\limits_{k=0}^{n-1}\mathcal{U}^*_{jk} \mathcal{U}_{j^{\prime}k} \right]
   \xrightarrow[\mathcal{U}]{\quad }\max
   \label{lagrangetovariateNUDlen}
\end{align}
we derive a novel algebraic problem\cite{malyshkin2019radonnikodym}
\begin{align}
  S \mathcal{U} &= \lambda \mathcal{U}
  \label{eigenvaluesLikeProblem}
\end{align}
The superoperator $S$ is a tensor
$S_{jk;j^{\prime}k^{\prime}}$, 
the ``eigenvector'' $\mathcal{U}$ is a partially unitary operator
$\mathcal{U}_{jk}$,
and the ``eigenvalue'' $\lambda$ is a Hermitian $D \times D$ matrix of Lagrange multipliers  $\lambda_{ij}$,
an ``eigenmatrix'' instead of the usual eigenvalue-scalar.
This is a new algebraic problem,
obtained as a Lagrange stationarity condition,
and currently, only a numerical solution is available.
Possible bounds on the maximum and minimum fidelity can be estimated by using,
instead of the full constraints (\ref{unitarityCond}),
the simplified one (\ref{unitarityCondSimplified}).
With the simplified constraints, the problem is equivalent to an eigenproblem;
with the full constraints (\ref{unitarityCond}),
it becomes the new algebraic problem (\ref{eigenvaluesLikeProblem}).
For a given tensor $S_{jk;j^{\prime}k^{\prime}}$ (\ref{Fidelity}), the (partially) unitary operator $\mathcal{U}_{jk}$
at the global maximum provides the best solution (maximal fidelity) to the quantum mechanics inverse problem.
The value of the fidelity equals the trace of the eigenmatrix,
$\mathcal{F}=\mathrm{Tr}\lambda$.

This optimization problem can be reduced to the algebraic problem (\ref{eigenvaluesLikeProblem})
only when the fidelity, $\mathcal{F}$, is a quadratic function (\ref{Fidelity}) of $\mathcal{U}$,
or, more generally, when it is of the Rayleigh-quotient form (\ref{fidelityProjectionsApproximationBBSProjectionsU}).
For other forms of $\mathcal{F}$, such as higher-order polynomials or expressions involving
the square root of multiple quadratic forms, the corresponding algebraic problem cannot be formulated.
In such cases, numerical algorithms involving general mathematical analysis tools and techniques ---
such as derivatives, gradients, gradient descent, and the Hessian matrix --- are typically employed \cite{wen2013feasible}.
Our numerical solution \cite{belov2024partiallyPRE,belov2024quantumPRE}
represents a transition  from relying on mathematical analysis tools to employing algebraic methods ---
specifically, eigenproblem-type techniques ---
where solving an eigenvalue problem serves as a building block of our iterative algorithm
to the problem (\ref{eigenvaluesLikeProblem}).
This algebraic approach is applicable only when the optimization problem is a QCQP.

\section{\label{BasicPropertiesOfSulU}Basic properties of the algebraic problem
$S \mathcal{U} = \lambda \mathcal{U}$}
The algebraic problem (\ref{eigenvaluesLikeProblem}) has a number of remarkable features.
It has a number of solutions
\begin{align}
  S \mathcal{U}^{[s]} &= \lambda^{[s]} \mathcal{U}^{[s]}
  \label{eigenvaluesLikeProblemSlev}
\end{align}
The total number of solutions is up to $Dn$, $s=0\dots Dn-1$.
For Hermitian tensor $S_{jk;j^{\prime}k^{\prime}}$ (\ref{Fidelity}),
the eigenmatrix  $\lambda^{[s]}$ is a Hermitian matrix of dimension $D\times D$.
The fidelity $\mathcal{F}$  (\ref{Fidelity}) in the state $\mathcal{U}^{[s]}$
is equal to the trace
$\mathcal{F}=\mathrm{Tr}\lambda^{[s]}$.
 Compare this with a regular eigenproblem, $\Ket{H|\psi}=\lambda \Ket{\psi}$,
 obtained from (\ref{FHamiltonian}), where $\lambda^{[s]}$ is a scalar and
$\mathcal{F}=\lambda^{[s]}$.
The superoperator $S_{jk;j^{\prime}k^{\prime}}$ can be interpreted as encompassing many Hamiltonians,
where each solution (\ref{eigenvaluesLikeProblemSlev}) corresponds to one of them.
The Lagrange multiplier matrix $\lambda^{[s]}$ can be interpreted as the ``Hamiltonian'' of a quantum system.
The ``ground state'' solution, selected from among these quantum systems,
corresponds to the solution with the maximum
$\mathrm{Tr} \lambda^{[s]}$.

These ``Hamiltonians'' $\lambda^{[s]}$ should be distinguished from regular Hamiltonians
that describe the quantum system dynamics,
which is given by the logarithm of $\mathcal{U}^{[s]}$ (\ref{logUCalc}).
In this way, the solution to the algebraic problem, represented by the pair  $\lambda^{[s]},\mathcal{U}^{[s]}$,
describes \textsl{two} Hamiltonians.
While the dynamics governed by $i\ln \mathcal{U}^{[s]}$ (\ref{logUCalc})
represent regular quantum state dynamics,
the dynamics governed by $\lambda^{[s]}$ can be interpreted as the evolution of the quantum channel itself,
possibly describing the time evolution of the quantum system as a whole.
This is our speculation that this ``second Hamiltonian'' $\lambda^{[s]}$
may possibly generate an evolution of the quantum system itself,
distinct from the regular explicit time dependence of the Hamiltonian $H(t)$.
The reason we consider this as ``second Hamiltonian'' is that,
for a simple form of $S_{jk;j^{\prime}k^{\prime}}$ such as the two-Hamiltonian approximation (\ref{TwoHamiltonianModel}),
the Lagrange multipliers $\lambda_{jj^{\prime}}$ and $\nu_{kk^{\prime}}$ describe a standard higher-order quantum map (\ref{evolution2}).
This leads us to conclude that a more general structure of $S_{jk;j^{\prime}k^{\prime}}$
could potentially describe a more general form of evolution of the quantum system itself.
If this ``second dynamics'' does not exist in nature,
then the meaning of $\lambda$ is reduced solely to determining the fidelity $\mathcal{F}=\mathrm{Tr}\lambda$
of the quantum inverse problem solution
and the only source of the quantum system's self-evolution is $H(t)$.
The $\lambda^{[s]}$ and $\mathcal{U}^{[s]}$ are related as described in Eq. (\ref{lambdaFromU}).

Orthogonality conditions are somewhat unusual.
If we consider Eq. (\ref{eigenvaluesLikeProblemSlev}) to hold for all components
without exception for certain projections (see Appendix \ref{SolutionsHierarchyConstraints} below),
we obtain the orthogonality condition.
By multiplying the equation for $s$ by $\mathcal{U}^{[s^{\prime}]}$ and summing, we derive
the orthogonality condition\footnote{
Eq. (\ref{otrCondition}) is written in superoperator notation.
Eq. (\ref{otrConditionExplicit}) is equivalent but presented in explicit form for the convenience of algorithm developers.
For clarity, a similar double-form representation is employed in several instances below ---
for example, in (\ref{vUTransfromNewBasis}) and (\ref{vUTransfromNewBasisMatrix}), among others.
Some expressions, such as (\ref{SBsisTransformed}) and (\ref{lambdaFromU}),
are given only in explicit form, as the superoperator notation may not be very clear.
}
\begin{align}
\Braket{\mathcal{U}^{[s^{\prime}]} | \lambda^{[s^{\prime}]\,\dagger}|\mathcal{U}^{[s]}}
&=
\Braket{\mathcal{U}^{[s^{\prime}]} | \lambda^{[s]} | \mathcal{U}^{[s]}}
\label{otrCondition} \\
\sum\limits_{j,j^{\prime}=0}^{D-1}\sum\limits_{k=0}^{n-1}
\left(\lambda_{jj^{\prime}}^{[s^{\prime}]} \mathcal{U}^{[s^{\prime}]}_{j^{\prime}k}\right)^* \mathcal{U}^{[s]}_{jk}
&=
\sum\limits_{j,j^{\prime}=0}^{D-1}\sum\limits_{k=0}^{n-1}
\mathcal{U}^{[s^{\prime}]*}_{jk}
\lambda_{jj^{\prime}}^{[s]}\mathcal{U}^{[s]}_{j^{\prime}k}
\label{otrConditionExplicit}
\end{align}
This is analogous to the familiar orthogonality condition for eigenvectors
 $\delta_{ss^{\prime}}=\Braket{\psi^{[s]}|\psi^{[s^{\prime}]}}$.
In traditional quantum mechanics, where $\lambda^{[s]}$ is a scalar
(which can be considered as $D=1$, $S_{0k;0q}=H_{kq}$),
the relation (\ref{otrCondition}) in the non-degenerate case $\lambda^{[s]}\ne\lambda^{[s^{\prime}]}$
leads to a $\lambda$-independent orthogonality condition $0=\Braket{\psi^{[s]}|\psi^{[s^{\prime}]}}$;
the zeros of the corresponding characteristic polynomial $0=\det \|H-\lambda \mathds{1}\|$ yield the eigenvalues $\lambda^{[s]}$.
In contrast, when $\lambda^{[s]}$ is a Hermitian matrix and
$\mathcal{U}^{[s]}$ is a unitary operator, the orthogonality condition (\ref{otrCondition}) depends on $\lambda^{[s]}$;
the equation for the eigenmatrices $\lambda^{[s]}$ cannot be formulated independently of the eigenstates $\mathcal{U}^{[s]}$,
unlike  the case of the characteristic polynomial in the standard eigenvalue problem (\ref{stationarySchrodinger}).
Note that in numerical methods ---
both for the regular eigenproblem \cite{anderson1999lapack}
and in our algorithm \cite{belov2024partiallyPRE,belov2024quantumPRE}
for the algebraic problem (\ref{eigenvaluesLikeProblem}) ---
each $s$-th (eigenvalue, eigenstate) pair is computed simultaneously;
there is no separation between finding the eigenvalue first and then determining the eigenstate.
The possibility of formulating a separate problem for eigenmatrices $\lambda$ that corresponds
to the algebraic problem (\ref{eigenvaluesLikeProblem})
and serves as an analogue of the zeros of the characteristic polynomial
in the eigenproblem (\ref{stationarySchrodinger}) is a subject of future research.

Consider the matrix of fidelity $\mathcal{F}_{ss^{\prime}}$ in the basis of the solutions 
$\mathcal{U}^{[s]}$ (\ref{eigenvaluesLikeProblemSlev}).
\begin{align}
\mathcal{F}_{ss^{\prime}}&=
\Braket{\mathcal{U}^{[s]}|S|\mathcal{U}^{[s^{\prime}]}}
=\sum\limits_{j,j^{\prime}=0}^{D-1}\sum\limits_{k,k^{\prime}=0}^{n-1}
\mathcal{U}^{[s]*}_{jk}
S_{jk;j^{\prime}k^{\prime}}
\mathcal{U}^{[s^\prime]}_{j^{\prime}k^{\prime}}
\label{Fss} 
\end{align}
One can directly verify that the orthogonality condition (\ref{otrCondition})
for Eq. (\ref{eigenvaluesLikeProblemSlev}) is consistent with the
Hermiticity
of
$\mathcal{F}_{ss^{\prime}}=\mathcal{F}^*_{s^{\prime}s}$.

The inconvenience of the previous state orthogonality condition (\ref{otrCondition})
lies in the difficulty of representing the tensor $S_{jk;j^{\prime}k^{\prime}}$
as a superposition of contributions from different
$\mathcal{U}^{[s]}_{jk}$ solutions,
as well as challenges in numerical implementation (see Appendix \ref{SolutionsHierarchyConstraints} below).
In \cite{belov2024quantumPRE}, we introduced a state orthogonality condition in the form 
\begin{align}
0&=\mathcal{F}_{ss^{\prime}}=\Braket{\mathcal{U}^{[s]} | S |\mathcal{U}^{[s^{\prime}]}}
\label{otrConditionS0} 
\end{align}
which corresponds to the matrix
$\mathcal{F}_{ss^{\prime}}$ (\ref{Fss}) being zero for $s\ne s^{\prime}$.
The condition (\ref{otrConditionS0}) allows for the tensor $S_{jk;j^{\prime}k^{\prime}}$
to be represented as
\begin{align}
S&\approx\sum\limits_{s=0}^{N_s-1} \frac{1}{\mathcal{F}_{ss}}
\Ket{S\middle|\mathcal{U}^{[s]}}
\Bra{\mathcal{U}^{[s]}\middle| S }
\label{expansionSinUs}
\end{align}
The main advantage of (\ref{otrConditionS0})
is that the matrix $\mathcal{F}_{ss^{\prime}}$ (\ref{Fss})
for an approximated $S$ from (\ref{expansionSinUs}) with the full $N_s=Dn$ basis
is equal to the corresponding matrix with the exact $S$.
The numerical implementation is also significantly simplified.
The main disadvantage of (\ref{otrConditionS0}) is that, for the solution $s$ in the hierarchy,
Eq. (\ref{eigenvaluesLikeProblemSlev})
is not satisfied for certain projections,
the number of which equals the number of previous states in the hierarchy.
However, the relation $\mathcal{F}_{ss}=\mathrm{Tr}\lambda^{[s]}$ always holds exactly.
This follows from the construction method of the solutions hierarchy:
we determine $\mathcal{U}^{[s]}_{jk}$  by maximizing $\mathcal{F}$  (\ref{Fidelity}),
with $\mathcal{U}^{[s]}_{jk}$ subject to the linear constraints imposed by the previous states
($s^{\prime}=0\dots s-1$), whether using (\ref{otrConditionS0}) or (\ref{otrCondition}).
This problem does not arise for a regular eigenproblem
($D=1$ and $\lambda^{[s]}$ is a scalar),
where the conditions (\ref{otrConditionS0}) and (\ref{otrCondition}) coincide.
For the ground state ($s=0$), all projections are always satisfied, as there are no prior state constraints applied.

A typical application of the (\ref{otrConditionS0}) hierarchy is that,
for a non-unitary operator $\mathcal{V}$ providing some fidelity $\mathcal{F}=\Braket{\mathcal{V}|S|\mathcal{V}}$,
we can construct a mixed unitary channel (\ref{UnitaryHierarchyQC}) that achieves the same fidelity.
\begin{align}
\mathcal{V}&=\sum\limits_{s=0}^{N_s-1} w_s \mathcal{U}^{[s]}
\label{VSumUs} \\
w_s&=
\frac{\Braket{\mathcal{V}|S|\mathcal{U}^{[s]}}}
{F_{ss}}
\label{wsExpr} \\
\Braket{\mathcal{V}|S|\mathcal{V}}&=
\sum\limits_{s=0}^{N_s-1} |w_s|^2 \Braket{\mathcal{U}^{[s]}|S|\mathcal{U}^{[s]}}
\label{qcAsSum}
\end{align}
Equation (\ref{qcAsSum}) holds exactly only for the full basis $N_s=Dn$.
Note that the Kraus rank (the minimal number of terms $N_s$ in (\ref{KrausOperator}))
may differ from the mixed unitary rank (the minimal number of terms $N_s$ in (\ref{UnitaryHierarchyQC})),
where $1=\sum_{s=0}^{N_s-1}|w_s|^2$.
\begin{align}
  A^{OUT}&=\sum\limits_{s=0}^{N_s-1} |w_s|^2 \mathcal{U}^{[s]} A^{IN} \mathcal{U}^{[s]\,\dagger}
  \label{UnitaryHierarchyQC} 
\end{align}

An important feature of a regular eigenproblem is the ability to find a unitary transformation that determines
the basis in which the matrix is diagonal. The problem we consider in this paper is more complex,
as there are multiple solutions to (\ref{eigenvaluesLikeProblem}),
and each solution consists of a pair of matrices, $\lambda^{[s]}_{ij}, \mathcal{U}^{[s]}_{jk}$.
The hierarchy we constructed above relates these different solutions.
Contrary to the regular eigenproblem, the selection of a linear constraint for orthogonality
to the previous solution can take different forms, such as (\ref{otrConditionS0}) or (\ref{otrCondition}).
Each form creates its own hierarchy.

\subsection{\label{densityOfStates}The density of states of the solutions hierarchy}
Let there be a hierarchy of solutions,  $\lambda^{[s]}_{ij}, \mathcal{U}^{[s]}_{jk}$,
and suppose we want to classify them according to some scalar criteria.
In traditional quantum mechanics, this classification is based on the Hamiltonian eigenvalues,
which determine the density of states --- that is, the number of eigenvalues within a given energy range.
Now, we have a superoperator $S_{jk;j^{\prime}k^{\prime}}$ instead of a Hamiltonian and eigenmatrices $\lambda^{[s]}_{ij}$
instead of energy levels. Since the fidelity of $\mathcal{U}^{[s]}_{jk}$ is equal to the trace of the eigenmatrix,
$\mathcal{F}=\mathrm{Tr}\lambda^{[s]}$,
a straightforward way to classify the states based on a scalar parameter is to classify them according to the solution's fidelity.
This ``density of states'' will describe not only the quantum system (\ref{logUCalc})
corresponding to $\mathcal{U}^{[s]}_{jk}$, but also the sample of $M$ observations (\ref{purePsiMap})
from which the data is taken to construct $S_{jk;j^{\prime}k^{\prime}}$.
If $S_{jk;j^{\prime}k^{\prime}}$  is obtained (\ref{SexamplePureExact}) from the states $\psi^{(t)}$
of a quantum system with Hamiltonian $H$ undergoing Schr\"{o}dinger equation (\ref{SchrodingerOriginal}) dynamics,
the ground state solution $\mathcal{U}^{[0]}_{jk}$ is precisely given by (\ref{Uquantum}),
and the total fidelity is  $\mathcal{F}^{[0]} = \mathrm{Tr}\lambda^{[0]} = M$.
But the algebraic problem (\ref{eigenvaluesLikeProblem}) has a number of other solutions,
$\mathcal{U}^{[s]}_{jk}$, each with fidelity $\mathcal{F}^{[s]}\le\mathcal{F}^{[0]}$.
The corresponding Hamiltonian $H^{[s]}$ can be obtained from $\mathcal{U}^{[s]}_{jk}$ as in (\ref{logUCalc}).
The question is how the $H^{[s]}$ corresponding to other solutions $\mathcal{U}^{[s]}_{jk}$
relate to the Hamiltonian $H$ used to generate the sample.
The first characteristic to consider is the density of states -- the number of solutions in the interval
$[\mathcal{F},\mathcal{F}+d\mathcal{F}]$.
From symmetry considerations, there should be a single solution
(unless the problem is degenerate, as in (\ref{TwoHamiltonianModel}))
corresponding to the minimal $\mathcal{F}$
(such minimal $\mathcal{F}$ solutions arise in the study of variational quantum algorithms\cite{cerezo2021variational,park2024hamiltonian}).
The density of states exhibits one or more maxima in the intermediate region, and we believe that this distribution
is primarily influenced by the sample of states (\ref{purePsiMap}) used to create $S_{jk;j^{\prime}k^{\prime}}$,
rather than by the Hamiltonian $H$ governing the underlying state evolution.
The current numerical implementation\cite{polynomialcode} has difficulty obtaining more than
a dozen solutions to (\ref{eigenvaluesLikeProblem}).
Thus, the study of the density of states will be the subject of our future research.

\subsection{\label{canonicFormBasis}Canonical form of a unitary quantum channel}
We considered above the hierarchy of solutions to the algebraic problem (\ref{eigenvaluesLikeProblem}).
Now, let us construct the transformation of a given solution into the basis in which it takes its canonical form.
Consider a solution $\lambda_{ij}, \mathcal{U}_{jk}$ of (\ref{eigenvaluesLikeProblem}).
The matrix $\lambda_{ij}$ is a Hermitian matrix corresponding to a ``second Hamiltonian'', and $\mathcal{U}_{jk}$
is a partially unitary operator that satisfies (\ref{unitarityCond})
and has the corresponding Hamiltonian given by (\ref{logUCalc}).
We define the canonical form of this solution as the one in which $\lambda_{ij}$
is diagonal and $\mathcal{U}_{jk}$ is the unit matrix.\\
\textbf{Theorem:} \textit{
Any $\lambda_{ij}, \mathcal{U}_{jk}$ solution can be transformed to a canonical basis when $D=n$.}\\
\textbf{Proof:}
The original $\mathcal{U}$ can be expressed as
\begin{align}
\mathcal{U}&=
\mathfrak{U}^{\dagger} \mathfrak{U} \mathcal{U} \mathfrak{V}^{\dagger}  \mathfrak{V}
\label{vUTransfrom}
\end{align}
where $\mathfrak{U}$ and $\mathfrak{V}$ are arbitrary unitary operators.
Then,
\begin{align}
\widetilde{\mathcal{U}}&=
\mathfrak{U} \mathcal{U} \mathfrak{V}^{\dagger}
\label{vUTransfromNewBasis} \\
\widetilde{\mathcal{U}}_{jk}&=
\sum\limits_{i=0}^{D-1}
\sum\limits_{q=0}^{n-1}\mathfrak{U}_{ji}\mathcal{U}_{iq}\mathfrak{V}^*_{kq}
\label{vUTransfromNewBasisMatrix}
\end{align}
can be considered as $\mathcal{U}$ in the new basis.
The transformation (\ref{vUTransfromNewBasis}) preserves unitarity.
The wrapping operators $\mathfrak{U}^{\dagger}$ and $\mathfrak{V}$
in (\ref{vUTransfrom})
are included in $S$ to obtain $\widetilde{S}$
in the new basis determined by chosen $\mathfrak{U}$ and $\mathfrak{V}$.
Thus, the simultaneous transformation of
$\mathcal{U}$ to $\widetilde{\mathcal{U}}$
and $S$ to $\widetilde{S}$ 
is an identity transformation (\ref{vUTransfrom});
it can be viewed as a change of representation in traditional quantum mechanics.
\begin{align}
\mathcal{F}&=\Braket{\mathcal{U}|S|\mathcal{U}}=
\Braket{\widetilde{\mathcal{U}}|\widetilde{S}|\widetilde{\mathcal{U}}}
\label{Ftransf} \\
\widetilde{S}_{jk;j^{\prime}k^{\prime}}&=
\sum\limits_{i,i^{\prime}=0}^{D-1}
\sum\limits_{q,q^{\prime}=0}^{n-1}
\mathfrak{U}_{ji}\mathfrak{V}^*_{kq}
S_{iq;i^{\prime}q^{\prime}}
\mathfrak{U}^*_{j^{\prime}i^{\prime}}\mathfrak{V}_{k^{\prime}q^{\prime}}
\label{SBsisTransformed}
\end{align}
For a given solution $\lambda_{ij},\mathcal{U}_{jk}$ to (\ref{eigenvaluesLikeProblem}),
the basis transformation to the canonical form consists of two steps:
\begin{itemize}
\item
Select $\mathfrak{U}$ such that $\lambda_{ij}$  becomes diagonal.
This corresponds to Eq. (B8) of Ref. \cite{belov2024quantumPRE},
where the unitary operator $\mathfrak{U}$ is constructed by containing all the eigenvectors of the matrix $\lambda_{ij}$.
This transformation is possible if $D\le n$.
\item
Select $\mathfrak{V}$ in such a way that $\widetilde{\mathcal{U}}$
(\ref{vUTransfromNewBasis}) becomes diagonal.
Choose
\begin{align}
\mathfrak{V}&=
\mathfrak{U}
\mathcal{U}
\label{Vtransf}
\end{align}
Here, $\mathcal{U}$ represents the solution in the initial basis,
and the operator $\mathfrak{U}$ is obtained in the previous step as the eigenvectors of $\lambda_{ij}$.
This transformation is possible only if $D=n$, as $\mathfrak{V}$ is unitary only in this case.
\end{itemize}
Thus, without loss of generality, any given stationary solution $\lambda_{ij}, \mathcal{U}_{jk}$
of (\ref{eigenvaluesLikeProblem})
can be considered with $\lambda_{ij}$ being diagonal and $\mathcal{U}_{jk}$
being the unit matrix.
Otherwise, the described transformations $\mathfrak{U}$ and $\mathfrak{V}$
should be applied as shown in Eqs. (\ref{vUTransfromNewBasis}) and (\ref{SBsisTransformed}).
$\blacksquare$
\begin{align}
\lambda_{ij}&=\lambda_i\delta_{ij}
\label{lambdaDiag} \\
\mathcal{U}_{jk}&=\delta_{jk} \label{Udiag}
\end{align}
\textbf{Note:} This algorithm transforms only a single given solution to canonical form.
We do not have a method for simultaneously converting several solutions to canonical form.
The reason we were able to simultaneously satisfy both (\ref{lambdaDiag}) and (\ref{Udiag})
is that for an arbitrary unitary $\mathfrak{V}$, the transformation $\widetilde{\mathcal{U}}= \mathcal{U} \mathfrak{V}$,
with the corresponding adjustment of $S$ (\ref{SBsisTransformed}), does not change the Lagrange multipliers $\lambda_{ij}$.
This allowed us to first choose $\mathfrak{U}$ to satisfy (\ref{lambdaDiag}) and then $\mathfrak{V}$ to satisfy (\ref{Udiag}).
See \texttt{\seqsplit{com/polytechnik/kgo/TestCanonicalForm.java}} for a unit test that demonstrates this theorem's proof.
Also note that since $\lambda_{ij}$ and $\mathcal{U}_{jk}$ are both diagonal in the canonical basis, they commute.
It might appear that $U\mathcal{U}$, where $U$ is an arbitrary diagonal unitary operator (which also commutes with $\lambda$),
could be a solution for the same $S$.
However, this is not the case because the tensor $S_{jk;j^{\prime}k^{\prime}}$
acts on $\mathcal{U}_{jk}$  as a whole,
i.e. in the canonical basis
$\sum_i S_{jk;ii}=\lambda_j \delta_{jk}$,
which means non-associative multiplication: $S (\mathcal{U}U)\ne (S \mathcal{U})U$.
Whereas traditional quantum mechanics deals with
non-commuting but always associative matrix multiplication\cite{born1926quantenmechanik},
our theory employs tensor multiplication,
which is associative only in special cases of $S$,
such as (\ref{groundStateApproximation}), or in certain situations (\ref{TwoHamiltonianModel}).
The distinction between tensor and matrix multiplication implies that, in the canonical basis,
$U\mathcal{U}$ is not a solution for a diagonal unitary operator $U$.
For a general non-degenerate $S_{jk;j^{\prime}k^{\prime}}$,
the only available degree of freedom is a common phase $\exp(i\xi)\mathcal{U}_{jk}$.

\section{\label{PostmeasurementStateDestruction}Measurement Unit and Post-Measurement State Collapse Rule}
In traditional quantum mechanics, any observable corresponds to a Hermitian operator $R$,
and its expected value is determined for a quantum state $\Ket{\psi}$ as
\begin{align}
R_{ex}&=\Braket{\psi|R|\psi}
\label{RmeasuredPsiPure}
\end{align}
This equation is often interpreted to mean that the actually observed value corresponds to one of the eigenvalues $R^{[s]}$ of the operator $R$,
and not to the expected value $R_{ex}$.
Different realizations (eigenvalues $R^{[s]}$)
have associated probabilities, and Eq. (\ref{RmeasuredPsiPure}) provides the expected value.
Importantly, the act of measurement destroys the state $\Ket{\psi}$:
the state $\Ket{\psi}$ is reduced to a single eigenstate of $R$ due to the measurement process, a so-called wavefunction collapse\cite{griffiths2019introduction,marinescu2011classical}.

Similarly, in superstate quantum mechanics, we have a Hermitian superoperator $R_{jk;j^{\prime}k^{\prime}}$
and its corresponding observable scalar value.
The expected value
$R_{ex}$
of the measurement process in the $\mathcal{U}_{jk}$ pure state and in the $\Upsilon$
mixed state (\ref{densMatrConvexSuperposFidelity}) is given by:
\begin{align}
R_{ex}&=\Braket{\mathcal{U}|R|\mathcal{U}}=\sum\limits_{j,j^{\prime}=0}^{D-1}\sum\limits_{k,k^{\prime}=0}^{n-1}
\mathcal{U}^*_{jk}
R_{jk;j^{\prime}k^{\prime}}
\mathcal{U}_{j^{\prime}k^{\prime}}
\label{RmeasuredPureU} \\
R_{ex}&=\mathrm{Tr} \Upsilon R=
\sum\limits_{s=0}^{N_s-1} P^{[s]} \Braket{\mathcal{U}^{[s]}|R|\mathcal{U}^{[s]}}
\label{RmeasuredMixedUpsilon}
\end{align}
In traditional quantum mechanics, the state $\Ket{\psi}$ is destroyed after the measurement act (\ref{RmeasuredPsiPure}).
In contrast, the measurement (\ref{RmeasuredPureU}) involves a quantum channel $\mathcal{U}$ as the state.
The measurement process may affect the wavefunction $\Ket{\psi}$ and the unitary quantum channel $\mathcal{U}$ differently --
the post-measurement destruction rules for wavefunctions and quantum channels may ultimately differ.

We were previously concerned mostly with the computational aspects of the quantum inverse problem. Let us speculate:
what would happen if a physical process capable of solving the quantum inverse problem in a single measurement existed in nature?
In traditional quantum mechanics,
the approach of a positive operator-valued measure (POVM) generalizes the familiar projection-valued measure \cite{nielsen2010quantum}.
In our theory, the eigenvalue itself can be a matrix. For example, consider the case where $R=S$, and
$\mathcal{U}^{[s]}$  and $\lambda^{[s]}$
are the solutions and corresponding eigenmatrices of (\ref{eigenvaluesLikeProblemSlev}).
A traditional measurement act can be interpreted as determining which corresponding operator eigenvalue is observed,
while the expected value is given by (\ref{RmeasuredPsiPure}).
Similarly, in superstate quantum mechanics, the measurement act can potentially be interpreted as determining which of its eigenmatrices
(when $R=S$, the ``second Hamiltonians'' $\lambda$ that are contained in $S$) are actually realized,
whereas the expected value of $\lambda$ in state $\mathcal{U}$
is given by the formula used in the numerical algorithm \cite{belov2024partiallyPRE},
which generalizes the expression $\lambda = \Braket{\psi|H|\psi}$ from the standard eigenvalue problem.
\begin{align}
\lambda_{ji}&=\mathrm{Herm}\sum\limits_{m=0}^{D-1}\sum\limits_{k,q=0}^{n-1}\mathcal{U}^*_{ik}S_{jk;mq}\mathcal{U}_{mq}
\label{lambdaFromU}
\end{align}
The expected scalar value is then $\lambda_{ex}=\mathrm{Tr} \lambda_{ij}$ (\ref{RmeasuredPureU}).
This raises the question of what the measurable unit is:
whether it is a scalar (as in traditional quantum mechanics)
or a Hermitian matrix? Since the corresponding algebraic problem (\ref{eigenvaluesLikeProblem})
has eigenmatrices as its spectrum, rather than the usual eigenvalues (scalars),
we are inclined to believe that the measurement unit is a Hermitian matrix.
For example, in traditional quantum mechanics,
the most commonly measured or calculated value is the minimal eigenvalue of the Hamiltonian,
the ground state energy level.
Similarly, in superstate quantum mechanics, the most important characteristic is the state corresponding
to the maximal fidelity (\ref{Fidelity}), which is obtained as a solution to the quantum inverse problem.
By solving this, we find the $\lambda^{[0]}$ that provides the maximal possible fidelity
$\mathcal{F}=\mathrm{Tr} \lambda^{[0]}$.
By analogy with quantum mechanics, if a physical process existed that could solve
the quantum inverse problem in a single measurement,
the object that could be directly observed would naturally be a Hermitian matrix.
One may also draw an analogy with a ``nondemolition'' measurement in traditional quantum mechanics:
If the measurement reports that the outcome is $\lambda^{[0]}$, it leaves the system in the state $\mathcal{U}^{[0]}$;
i.e.,
when the measurement yields $\lambda^{[0]}$,
it creates a quantum system with the regular Hamiltonian $i\ln \mathcal{U}^{[0]}$ (\ref{logUCalc}).
The matrices $\lambda^{[s]}$ each correspond to their own solution of the quantum inverse problem,
and they cannot be obtained using any set of
\href{https://en.wikipedia.org/wiki/POVM}{a positive operator-valued measure (POVM)}
operators\cite{chiribella2009theoretical}.

However, we currently remain in the domain of computation on classical computers.
For a given tensor $S_{jk;j^{\prime}k^{\prime}}$, 
we have a numerical algorithm that computes the pair $\lambda^{[0]}_{ij},\mathcal{U}^{[0]}_{jk}$.
This is analogous to calculating, from a given Hamiltonian, the eigenvalue-eigenvector pair
$\lambda^{[0]},\psi^{[0]}_j$,
corresponding to the ground state in traditional quantum mechanics.

\section{\label{Schrodinger}A time-dependent Schr\"{o}dinger-like equation for the dynamics of a unitary operator}

After deriving the stationary Schr\"{o}dinger-like equation (\ref{eigenvaluesLikeProblem}),
the next objective is to develop an analogue of the non-stationary Schr\"{o}dinger-like
equation to describe the dynamics of a quantum channel defined by a unitary operator.
The approach is actually a form of higher-order quantum maps
\cite{chiribella2008transforming,bisio2019theoretical,odake2024higher,taranto2025higher}.
Our paper is primarily concerned with applications to machine learning
and the potential for simulating quantum dynamics on classical computers.
For this reason, it is convenient to treat different quantum objects ---
wavefunctions, unitary operators, and quantum channel Kraus operators ---
in a unified way, as elements of a Hilbert space, each subject to its own constraints.
These constraints may include wavefunction normalization (\ref{statePsi}),
(partial) unitarity conditions (\ref{unitarityCond}),
and Kraus operator constraints (\ref{constraintKraussSpur});
all of these constraints are quadratic.
While formulating the minimal dynamics of a system, in addition to the traditional linear maps used
in higher-order quantum theory, we may also consider a simple form of nonlinear map in the form of a
\href{https://en.wikipedia.org/wiki/Gross\%E2\%80\%93Pitaevskii_equation}{Gross--Pitaevskii}-type
equation (GPE)\cite{gross1961structure,pitaevskii1961vortex},
which often arises in various quantum systems.

Whereas the stationary problem has a clear meaning in the context of the quantum inverse problem,
the non-stationary problem does not.
The central question is whether a physical process exists that can implement such dynamics.
This can be viewed as an alternative to using a time-dependent Hamiltonian
for constructing a time-dependent quantum channel.
In the field of computer modeling, this issue is of much less concern,
as many different transformations can be valuable in machine learning.
In this paper, we focus on formulating such dynamics based on frameworks with multiple quadratic constraints,
as this provides a solid foundation for various numerical implementations.
In \cite{belov2024quantumPRE}, we examined the equation
\begin{align}
i\hbar\frac{\partial \mathcal{U}}{\partial t}&= S \mathcal{U}
\label{SchrodingerNonStationary}
\end{align}
The solution becomes a unitary operator $\mathcal{U}$ instead of a vector
$\psi$, and the Hamiltonian $H$  is replaced by a superoperator $S$ (\ref{Fidelity}).
This approach allows all ``higher-order'' dynamics to be incorporated into a general Hermitian tensor $S_{jk;j^{\prime}k^{\prime}}$,
which is obtained from system observations, as discussed above.
In the stationary case, this tensor defines the quantum inverse problem.
In the non-stationary case, one can formally attempt to express the solution 
$\mathcal{U}^{(t)}$ as
\begin{align}
 \mathcal{U}^{(t)}&=
  \exp \left[-i\frac{t}{\hbar} S \right] \mathcal{U}^{(0)} \label{UquantumS}
\end{align}
by considering the tensor $S_{jk;j^{\prime}k^{\prime}}$
as a Hermitian matrix of dimensions $Dn\times Dn$
and $\mathcal{U}$ as a vector of dimension $Dn$.
However, the solution $\mathcal{U}^{(t)}$ (\ref{UquantumS})
only satisfies the simplified constraint (\ref{unitarityCondSimplified})
and fails to satisfy the unitarity conditions (\ref{unitarityCond}).\footnote{
The exponent of the Hermitian matrix in (\ref{Uquantum}) and (\ref{UquantumS})
is typically calculated as a Taylor series.
When the system's evolution needs to be obtained for a short time $\tau$,
a Crank-Nicolson\cite{crank1947practical,iitaka1994solving} step,
in which the exponential function is approximated by the
\href{https://en.wikipedia.org/wiki/Cayley_transform}{Cayley transform}, can be applied:
$\psi(t+\tau)=\left(1-\frac{i\tau}{2\hbar}H\right)\left(1+\frac{i\tau}{2\hbar}H\right)^{-1} \psi(t)+
O\left(\left(\frac{\tau}{\hbar} H\right)^3\right)$,
which is also a unitary transformation.
If full diagonalization of $H$ is feasible, the best approach to solving the time-dependent
Schr\"{o}dinger equation is to use (\ref{Uquantum}) in the basis of eigenvectors of $H$;
this approach provides the exact solution for any arbitrary $\tau$.
}

A general $S$, which encompasses many different Hamiltonians, complicates the problem.
However, for a special form of $S$, the situation simplifies significantly.
In \cite{belov2024quantumPRE}, we introduced a form containing only a single Hamiltonian, $\lambda$,
typically corresponding to the ground state, and referred to this $S$ as the single Hamiltonian approximation.
\begin{align}
S_{jk;j^{\prime}k^{\prime}}&\approx\lambda_{jj^{\prime}}\delta_{kk^{\prime}}
\label{groundStateApproximation}
\end{align}
This is a highly degenerate $S_{jk;j^{\prime}k^{\prime}}$.
The maximal fidelity then applies to an arbitrary unitary operator.
This form of $S$ is straightforward to consider and serves as a good starting point for various approximations,
but it lacks a key feature of the inverse problem ---
the existence of multiple solutions to the stationary problem (\ref{eigenvaluesLikeProblemSlev}),
where each eigenmatrix $\lambda^{[s]}$ corresponds to a distinct quantum system with its own fidelity,
$\mathcal{F}=\mathrm{Tr} \lambda^{[s]}$.

Another distinction from traditional quantum mechanics lies in the role of superpositions of solutions.
If the initial state $\mathcal{U}^{(0)}$ is a superposition of the solutions $\mathcal{U}^{[s]}$
to the stationary problem (\ref{eigenvaluesLikeProblemSlev}),
\begin{align}
\mathcal{U}^{(0)}&=\sum\limits_{s=0}^{N_s-1} a_s \mathcal{U}^{[s]}
\label{initialState}
\end{align}
then the initial $\mathcal{U}^{(0)}$ is not necessarily unitary.
The violation of unitarity arising from the state superposition
has deep physical significance.
A superposition of eigenstates (\ref{initialState}) is not a unitary operator and,
therefore, does not represent a physical state.
Two possible resolutions exist.

First one involves introducing the density supermatrix
$\Upsilon$.
 Similarly to how, in quantum mechanics, the density matrix is formed as a convex combination of pure states,
we can consider a convex combination of unitary channels as a mixed state in superstate quantum mechanics.
Note that not every quantum channel (\ref{KrausOperator}) can be represented as a mixed unitary one\cite{girard2022mixed}.
\begin{align}
\Upsilon&=\sum\limits_{s=0}^{N_s-1} P^{[s]} \Ket{\mathcal{U}^{[s]}} \Bra{\mathcal{U}^{[s]}}
\label{densMatrConvexSuperpos} \\
\mathcal{F}&=\mathrm{Tr} \Upsilon S=\sum\limits_{s=0}^{N_s-1} P^{[s]} \Braket{\mathcal{U}^{[s]}|S|\mathcal{U}^{[s]}}
\label{densMatrConvexSuperposFidelity}
\end{align}
The dynamics in traditional quantum mechanics (\ref{unitaryPsiEvolution}) transform a pure state into another pure state.
To model the time evolution of a quantum system that converts a pure state into a mixed state,
a common approach is to introduce additional terms into the Schr\"{o}dinger equation,
as exemplified by the \href{https://en.wikipedia.org/wiki/Lindbladian}{GKSL} (Gorini-Kossakowski-Sudarshan-Lindblad) equation.
The issue with a general $S_{jk;j^{\prime}k^{\prime}}$ is that the dynamics associated with it transforms
a pure state into a mixed one without introducing any additional terms in the dynamic equation.
Even if the initial state $\mathcal{U}^{(0)}$ is taken as the ground state $\mathcal{U}^{[0]}$,
it may still evolve into a mixed state under (\ref{SchrodingerNonStationary})-type dynamics.
The single-Hamiltonian approach (\ref{groundStateApproximation})
does not have this issue, but it is too simplistic to accurately describe a real quantum channel.
The consideration of high order linear maps lays in the concept
of higher-order quantum theory and attracts attention recently\cite{taranto2025higher}.
These maps are typically studied as abstract transformations, without explicitly involving
a quadratic ``energy''-like functional that defines the mapping.
An important result of our study is the introduction of the Hermitian tensor $S_{jk;j^{\prime}k^{\prime}}$,
obtained from system observations, which encapsulates the higher-order dynamics.
For a general $S_{jk;j^{\prime}k^{\prime}}$, the dynamics in the form we introduced in (\ref{SchrodingerNonStationary})
can be directly studied numerically and exhibit rich behavior, such as transforming a unitary mapping into a mixed-unitary channel.
Below, we consider its simplest case for $S$ in the two-Hamiltonian approximation,
which preserves unitarity and can be represented as a 2D graphical diagram in (\ref{qcExampleU}).

The second option, which cannot be reduced to higher-order quantum maps,
is to consider a non-linear equation for $\mathcal{U}$.
Since the superposition of solutions no longer plays the same role as in traditional quantum mechanics,
the dynamic equation is no longer required to be first-order.
The simplistic form that correctly satisfies the dynamics for any stationary solution is
\begin{align}
i\hbar\frac{\partial \mathcal{U}}{\partial t}&= \Braket{\mathcal{U}|S|\mathcal{U}} \mathcal{U}
\label{NonLinearSynamicsNonStationary}
\end{align}
It has the solution that any unitary operator evolves as 
\begin{align}
 \mathcal{U}^{(t)}&=\exp\left[-i\frac{t}{\hbar}\mathcal{F}\right]\mathcal{U}^{(0)}
 \label{solutionNonLinearSynamicsNonStationary}
\end{align}
with only the common exponent determined by the channel fidelity $\mathcal{F}=\Braket{\mathcal{U}|S|\mathcal{U}}$.
This correctly describes the dynamics of any stationary solution (\ref{eigenvaluesLikeProblemSlev}),
but for a general $\mathcal{U}^{(0)}$,
it results in rather simplistic dynamics.

The two most promising candidates for the dynamic equations are
(\ref{SchrodingerNonStationary}) and (\ref{NonLinearSynamicsNonStationary}).
One can also consider a combination of these equations with some weight coefficients, $a$ and $b$,
to obtain a Gross--Pitaevskii type
equation (GPE)\cite{gross1961structure,pitaevskii1961vortex}.
\begin{align}
i\hbar\frac{\partial \mathcal{U}}{\partial t}&= a S \mathcal{U}
+b\Braket{\mathcal{U}|S|\mathcal{U}}\mathcal{U}
\label{GrossPitaevsky}
\end{align}
Depending on the coefficients $a$ and $b$,
the equation can exhibit a rich variety of behavioral types and can thus be applied to the dynamics of various inverse problems,
whereas the original GPE describes the dynamics of various direct problems.

The most direct way to obtain the dynamic equation is to postulate the system's action
(typically in the form of an integral of the Lagrangian)
and then derive the dynamics from the
\href{https://en.wikipedia.org/wiki/Action_principles}{principle of least action};
see, for example, Ref. \cite{toptygin2013foundations}, problem 4.124 on page 318.
However, formulating a Lagrangian for a system can be challenging,
particularly when the dynamic equation is nonlinear, such as in the case of the GPE.
Currently, we do not have an expression for the Lagrangian from which the dynamic equation can be derived.
This remains a topic for future research.
The reasons we consider $S_{jk;j^{\prime}k^{\prime}}$ as a possible source of evolution of a quantum system itself are:
\begin{itemize}
\item Some specific forms of $S_{jk;j^{\prime}k^{\prime}}$ produce exactly higher-order quantum dynamics.
For example the (\ref{TwoHamiltonianModel}) model for $S_{jk;j^{\prime}k^{\prime}}$
describe a standard higher-order quantum map (\ref{evolution2}).
\item The equation (\ref{SchrodingerNonStationary}) closely resembles the Schr\"{o}dinger equation.
Similarly to traditional quantum mechanics, where the Hamiltonian $H$ governs both the stationary problem
and the non-stationary dynamics,
we may assume that $S_{jk;j^{\prime}k^{\prime}}$ can likewise govern both the stationary (quantum inverse problem)
and non-stationary (higher-order dynamics, or ``dynamics of dynamics'') cases.
\item This type of system evolution may be beneficial in machine learning and artificial intelligence, where we are not constrained by the limitations of physical-world processes.
\end{itemize}

Even without exact knowledge of the dynamic equation, we can proceed with constructing the computational model.
In traditional quantum computation,
researchers typically represent a quantum circuit as a sequence of unitary transformations (\ref{unitaryPsiEvolution})
rather than analyzing the corresponding dynamics (\ref{Uquantum}) governed by a given Hamiltonian.
For example, a simple quantum circuit transforming the initially prepared state
$\Ket{\psi^{(0)}}$ can take a form like this\footnote{
This is a just an example, the circuit is not intended to be functional.}
\begin{equation}
\hbox{
\begin{circuitikz}[line width=1pt]
\tikzstyle{operator} = [draw,fill=white,minimum size=1.5em] 
\tikzstyle{phase} = [draw,fill,shape=circle,minimum size=5pt,inner sep=0pt]
\tikzstyle{Xcross} = [path picture={ 
\draw[thick,black,inner sep=0pt]
(path picture bounding box.south east) -- (path picture bounding box.north west) (path picture bounding box.south west) -- (path picture bounding box.north east);
}]
\tikzstyle{cross} = [path picture={ 
\draw[thick,black](path picture bounding box.north) -- (path picture bounding box.south) (path picture bounding box.west) -- (path picture bounding box.east);
}]
\tikzstyle{Ocross} = [draw,circle,cross,minimum width=0.3 cm]

\ctikzset{multipoles/thickness=2}
\ctikzset{multipoles/dipchip/width=0.9}
\ctikzset{multipoles/external pins thickness=2}
\ctikzset{multipoles/dipchip/pin spacing=0.6}
\draw (0,0) node[dipchip,
line width=0.6pt,
num pins=10, no topmark,
external pins width=0.1,
hide numbers
,draw only pins={6,7,8,9,10}
](C){{\Large $\Ket{\psi^{(0)}}$}};

\node [above, font=\small] at (C.pin 10) {$0$};
\node [above, font=\small] at (C.pin 9) {$1$};
\node [above, font=\small] at (C.pin 8) {$2$};
\node [above, font=\small] at (C.pin 7) {$3$};
\node [above, font=\small] at (C.pin 6) {$4$};

\draw (C.pin 10) to[short,-] ++(0.8,0)
node[operator] (H1) {H}
to[short,-] ++(0.8,0)
node[Xcross] (H2) { }
to[short,-] ++(0.8,0)
node[operator] (H3) {H}
to[short, -] ++(0.8,0) node[] (nend) {};

\draw (C.pin 9) -- (C.pin 9 -| H3)
node[phase] (Pp2a) {} -- (C.pin 9 -| nend);

\draw (C.pin 8) -- (C.pin 8 -| H1)
node[operator] {H} -- (C.pin 8 -| H2)
node[Xcross]  {} -- (C.pin 8 -| H3)
node[phase] {} -- (C.pin 8 -| nend);

\draw (C.pin 7) -- (C.pin 7 -| H2)
node[phase] (Fr12) {} -- (C.pin 7 -| nend);

\draw (C.pin 6) -- (C.pin 6 -| H1)
node[operator] {S} -- (C.pin 6 -| H3)
node[Ocross] (PP2b) {} -- (C.pin 6 -| nend);

\draw (Pp2a.center) -- (PP2b.center);
\draw (Fr12.center) -- (H2.center);

\end{circuitikz}
}
\label{qcExample}
\end{equation}
Here, the $0\dots 4$ ``terminals'' of $\Ket{\psi^{(0)}}$ can take different forms.
The most common approach is to consider them as qubits,
with the state itself regarded as a direct (tensor) product of these qubits.
In (\ref{qcExample}), the state is a product of $m=5$ qubits.
One can vectorize this state to convert a state of tensor product of $m$ qubits
to a vector of $2^m$ complex coefficients.\footnote{
\label{vectorizedSpaceFN}
Note that in \cite{belov2024partiallyPRE,belov2024quantumPRE},
we represented $\Ket{\psi}$ and $\mathcal{U}$ in vectorized form
to simplify the integration of our numerical algorithm with numerical libraries\cite{anderson1999lapack}.
}
This exponential dependence imposes a natural limit on the size of a quantum system that can be modeled on a classical computer.
Assuming that the state of a single qubit is stored using 8 bytes of classical memory,
an $m=30$ qubit system requires $8\cdot 2^{30}$ bytes, i.e.,  $8,\mathrm{GiB}$ of memory.
A quantum system with $m=30$ qubits therefore represents a natural upper limit for practical consideration on a classical computer.

Quantum computation is a unitary transformation of a prepared state $\Ket{\psi^{(0)}}$ using a quantum circuit.
A number of unitary transformations proceed from left to right, and the circuit is represented as a 1D plot of sequential unitary transformations.
The combined unitary operator $\mathcal{U}$ corresponding to the
entire quantum circuit is related to the quantum system Hamiltonian
as shown in (\ref{logUCalc}).
The practical difficulties lie in the creation of both the quantum system and the initial prepared state required
to calculate the result of the unitary transformation.
If, instead of general computation, we limit our goal to determining the Hamiltonian ground state,
the problem becomes easier to approach.
Techniques such as variational quantum algorithms\cite{cerezo2021variational} and adiabatic quantum
computation\cite{albash2018adiabatic,aharonov2008adiabatic} can then be applied.

For higher-order dynamics, where the state is described not by a unit vector but by a unitary operator $\mathcal{U}$,
consider the transformations that preserve the unitarity of $\mathcal{U}$ (partial unitarity if $D<n$).
The transformation of a unitary operator $\mathcal{U}$ that preserves the constraints (\ref{unitarityCond}) is
\begin{align}
\widetilde{\mathcal{U}}&=
\mathfrak{U} \mathcal{U} \mathfrak{V}^{\dagger}
\label{transfUDynamics}
\end{align}
where $\mathfrak{U}$ is an arbitrary unitary operator of dimension $D\times D$,
and $\mathfrak{V}$ is an arbitrary unitary operator of dimension $n\times n$.
It is actually a transformation similar to (\ref{vUTransfromNewBasis}) that we considered above
in the canonical form study.
However, it is no longer an identity for $\mathcal{F}$ since there is no corresponding transformation of $S$  to compensate for the transformation of $\mathcal{U}$.

If the constraint (\ref{unitarityCond}) is not satisfied for the original state $\mathcal{U}$,
then the Gram matrix
is not a unit matrix, and it is straightforward to write a transformation
that preserves the Gram matrix (\ref{GramMatrix}).
However, for the time evolution of the system state $\mathcal{U}^{(t)}$,
we will consider only states that satisfy the unitarity constraints (\ref{unitarityCond}).
Thus, only transformations of the form (\ref{transfUDynamics}) should be allowed.

The transformation (\ref{transfUDynamics}) indicates that any possible time evolution of
$\mathcal{U}_{jk}$ can involve \textsl{two}
distinct sequences of unitary transformations:
$U$ (horizontal) and $V$ (verical).
\begin{align}
\mathcal{U}^{(t)}&=
U^{(t)}\dots U^{(\tau)} \mathcal{U}^{(0)} V^{(\tau)\dagger}\dots V^{(t)\dagger} \label{UtTimeEvolution}
\end{align}
In contrast, traditional quantum mechanics describes time evolution
using a single sequence of unitary operators $U$
\begin{align}
   \Ket{\psi^{(t)}}&=U^{(t)}\dots U^{(\tau)} \Ket{\psi^{(0)}} \label{PsitTimeEvolution}
\end{align}
where $U^{(t)}$ describes the evolution of the quantum system with Hamiltonian $H$ between
$t-\tau$ and $t$,
and is obtained from (\ref{Uquantum}).

The transformation that preserves unitarity is given by (\ref{UtTimeEvolution}),
but it is difficult to explicitly express the operators $U$ and $V$ from a general tensor $S_{jk;j^{\prime}k^{\prime}}$,
in a manner analogous to the relation (\ref{Uquantum}) between the evolution operator in (\ref{PsitTimeEvolution}) and the Hamiltonian.
However, if we consider a simplified form of $S_{jk;j^{\prime}k^{\prime}}$,
which we shall refer to as the two-Hamiltonian approximation,
\begin{align}
S_{jk;j^{\prime}k^{\prime}}&\approx\lambda_{jj^{\prime}}\delta_{kk^{\prime}} + \delta_{jj^{\prime}} \nu_{kk^{\prime}}
\label{TwoHamiltonianModel}
\end{align}
then the stationary equation becomes
\begin{align}
 S \mathcal{U} &= \lambda \mathcal{U} + \mathcal{U} \nu 
  \label{eigenvaluesLikeProblemUV}
\end{align}
where both $\lambda$ and $\nu$ are Hermitian matrices.
If the system dynamics is described by Eq. (\ref{SchrodingerNonStationary}),
then a time-dependent solution in the two-Hamiltonian approximation is
\begin{align}
\mathcal{U}^{(t)}&=
 \exp\left[-i\frac{t}{\hbar}\lambda\right] \mathcal{U}^{(0)}
 \exp\left[-i\frac{t}{\hbar}\nu\right]
 \label{evolution2}
\end{align}
It generates both the ``horizontal'' $U$ and ``vertical'' $V$ operators in (\ref{UtTimeEvolution}),
similar to the transformation in (\ref{PsitTimeEvolution}) with unitary evolution operator from Eq. (\ref{Uquantum}).
Equation (\ref{evolution2}) represents a higher-order quantum map \cite{taranto2025higher}.
This is an important result, showing that certain specific forms of $S_{jk;j^{\prime}k^{\prime}}$, such as (\ref{eigenvaluesLikeProblemUV}),
reproduce established results. This observation encourages us to relate $S_{jk;j^{\prime}k^{\prime}}$
not only to the stationary problem but also to higher-order dynamics.
A more general form of $S_{jk;j^{\prime}k^{\prime}}$ could potentially provide a rich class of mappings.
One remaining question concerns the origin of the Lagrange multipliers $\nu$.
The conditions (\ref{unitarityCond}) and (\ref{equivUnitaryCond}) are identical for $D = n$,
and in this case, the second set of constraints must be obtained from somewhere else.
The maximal fidelity, $\mathcal{F}=\mathrm{Tr}\lambda + \mathrm{Tr}\nu$,
applies to an arbitrary unitary operator,
the problem is degenerate.

Similarly to the circuit (\ref{qcExample}) for the wavefunction $\Ket{\psi^{(0)}}$ transformation,
one can construct a circuit for the transformation of the unitary operator $\mathcal{U}^{(0)}$.
This new circuit now involves time along both the horizontal and vertical axes.
The horizontally directed gates define a sequence of unitary gates $U^{(l)}$,
while the vertically directed gates define a sequence of unitary gates $V^{(l)}$.
These $U^{(l)}$ and $V^{(l)}$ can be arbitrary unitary gates, as in (\ref{UtTimeEvolution}),
and are not necessarily in the simplistic form (\ref{evolution2}) of the two-Hamiltonian approximation.
For a general 2D circuit, the $\mathcal{U}$-terminals may also be understood as qubits,
similar to those in (\ref{qcExample}) for the state $\Ket{\psi^{(0)}}$.
The figure below demonstrates an example of the unitary transformations $U$ and $V$ in Eq. (\ref{UtTimeEvolution}),
presented in a circuit-style 2D graphical representation.
The combined $U$ and $V$ represent two sequences of quantum gates,
analogous to a single sequence (\ref{PsitTimeEvolution}) in 1D quantum computations.

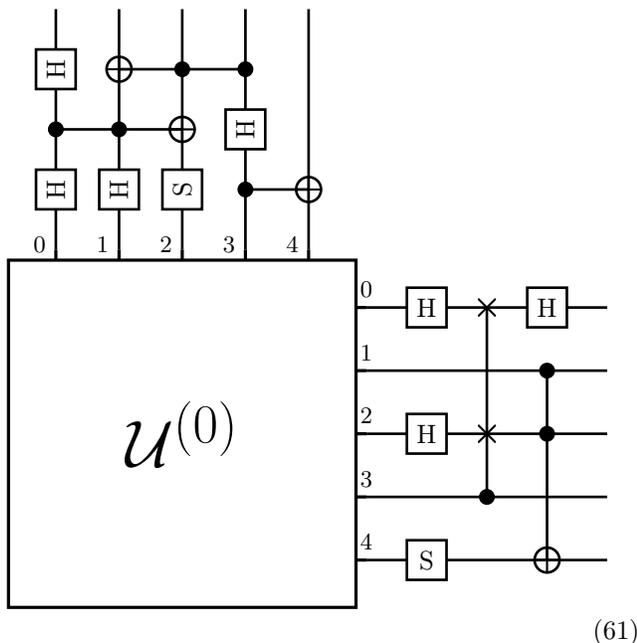
\begin{equation}
\hbox{
\begin{circuitikz}[line width=1pt]
\tikzstyle{operator} = [draw,fill=white,minimum size=1.5em] 
\tikzstyle{phase} = [draw,fill,shape=circle,minimum size=5pt,inner sep=0pt]
\tikzstyle{Xcross} = [path picture={ 
\draw[thick,black,inner sep=0pt]
(path picture bounding box.south east) -- (path picture bounding box.north west) (path picture bounding box.south west) -- (path picture bounding box.north east);
}]
\tikzstyle{cross} = [path picture={ 
\draw[thick,black](path picture bounding box.north) -- (path picture bounding box.south) (path picture bounding box.west) -- (path picture bounding box.east);
}]
\tikzstyle{Ocross} = [draw,circle,cross,minimum width=0.3 cm]

\ctikzset{multipoles/thickness=2}
\ctikzset{multipoles/external pins thickness=2}
\ctikzset{multipoles/qfpchip/pin spacing=0.6}

\node [qfpchip,
line width=0.6pt,
num pins=20, no topmark,
external pins width=0.1,
hide numbers,
draw only pins={11-20}
](C){{\Huge $\mathcal{U}^{(0)}$}};

\node [above, font=\small] at (C.pin 15) {$0$};
\node [above, font=\small] at (C.pin 14) {$1$};
\node [above, font=\small] at (C.pin 13) {$2$};
\node [above, font=\small] at (C.pin 12) {$3$};
\node [above, font=\small] at (C.pin 11) {$4$};

\draw (C.pin 15) to[short,-] ++(0.8,0)
node[operator] (H1) {H}
to[short,-] ++(0.8,0)
node[Xcross] (H2) { }
to[short,-] ++(0.8,0)
node[operator] (H3) {H}
to[short, -] ++(0.8,0) node[] (nend) {};

\draw (C.pin 14) -- (C.pin 14 -| H3)
node[phase] (Pp2a) {} -- (C.pin 14 -| nend);

\draw (C.pin 13) -- (C.pin 13 -| H1)
node[operator] {H} -- (C.pin 13 -| H2)
node[Xcross]  {} -- (C.pin 13 -| H3)
node[phase] {} -- (C.pin 13 -| nend);

\draw (C.pin 12) -- (C.pin 12 -| H2)
node[phase] (Fr12) {} -- (C.pin 12 -| nend);

\draw (C.pin 11) -- (C.pin 11 -| H1)
node[operator] {S} -- (C.pin 11 -| H3)
node[Ocross] (PP2b) {} -- (C.pin 11 -| nend);

\draw (Pp2a.center) -- (PP2b.center);
\draw (Fr12.center) -- (H2.center);

\node [left,yshift=2, font=\small] at (C.pin 20) {$0$};
\node [left,yshift=2, font=\small] at (C.pin 19) {$1$};
\node [left,yshift=2, font=\small] at (C.pin 18) {$2$};
\node [left,yshift=2, font=\small] at (C.pin 17) {$3$};
\node [left,yshift=2, font=\small] at (C.pin 16) {$4$};

\draw (C.pin 20) -- ++(0,0.8)
node[operator,rotate=90] (G1) {H} --  ++(0,0.8)
node[phase] (G2) {} -- ++(0,0.8)
node[operator,rotate=90] (G3) {H} --  ++(0,0.8)
node[] (nYend) {};

\draw (C.pin 19) -- (C.pin 19 |- G1)
node[operator,rotate=90] {H} --  (C.pin 19 |- G2)
node[phase]  {} -- (C.pin 19 |- G3)
node[Ocross] (oC3b) {} -- (C.pin 19 |- nYend);

\draw (C.pin 18) -- (C.pin 18 |- G1)
node[operator,rotate=90] {S} --  (C.pin 18 |- G2)
node[Ocross] (G2a) {} -- (C.pin 18 |- G3)
node[phase] (oC3) {} -- (C.pin 18 |- nYend);

\draw (G2.center) -- (G2a.center);

\draw (C.pin 17) -- (C.pin 17 |- G1)
node[phase] (oC1a) {} -- (C.pin 17 |- G2)
node[operator,rotate=90] {H} --  (C.pin 17 |- G3)
node[phase] (oC3a) {} -- (C.pin 17 |- nYend);

\draw (oC3b.center) -- (oC3a.center);

\draw (C.pin 16) -- (C.pin 16 |- G1)
node[Ocross] (oC1b) {} -- (C.pin 16 |- nYend);

\draw (oC1b.center) -- (oC1a.center);

\end{circuitikz}
}
\label{qcExampleU}
\end{equation}
Whereas a 1D circuit (\ref{qcExample}) converts a quantum state
$\Ket{\psi^{(0)}}$
into another quantum state $\Ket{\psi}$,
the 2D circuit (\ref{qcExampleU}) converts a quantum system $\mathcal{U}^{(0)}$
into another quantum system $\mathcal{U}$.\footnote{
The term 2D means that the circuit (\ref{qcExampleU}) transforms a 2D object $\mathcal{U}^{(0)}_{jk}$
into another 2D object,
as described by Eq. (\ref{UtTimeEvolution}).
This process generates two independent sequences of unitary transformations: a ``horizontal'' sequence $U^{(t)}$
and a ``vertical'' sequence $V^{(t)}$.
In contrast, standard 1D quantum evolution (\ref{qcExample}) transforms a 1D vector $\Ket{\psi^{(0)}}$
into another 1D vector,
according to Eq. (\ref{PsitTimeEvolution}), involving only a  ``horizontal'' sequence $U^{(t)}$.
Refer to Sec. \ref{examples2D} for illustrative examples.
}
This raises a question about the selection of the initial quantum system $\mathcal{U}^{(0)}$.
In regular quantum computations, the known low-entropy initial state $\Ket{\psi^{(0)}}$
is often considered the system's ground state, typically prepared using a cooling technique such as natural or laser cooling.
Similarly, in superstate quantum computations, the ``ground state'' $\mathcal{U}^{[0]}$ can be used as the initial state $\mathcal{U}^{(0)}$.
The result of the 2D computation is a new quantum system $\mathcal{U}$ (\ref{logUCalc}),
and the outcome of the computation is obtained by performing a measurement on the operator $\mathcal{U}$.
Note that the post-measurement destruction rules for wavefunctions and quantum systems likely differ,
as the measurement of a quantum state and the identification of a quantum system are two fundamentally different problems.

The unitary operator corresponding to the 1D circuit (\ref{qcExample})
directly defines a quantum system with the Hamiltonian (\ref{logUCalc}).
The 2D circuit (\ref{qcExampleU}) defines the combined horizontal $U$ and vertical $V$ unitary operators,
which can similarly be mapped to $\lambda$ and $\nu$, defining the higher-order map (\ref{evolution2}).
This map corresponds to the specific form of $S_{jk;j^{\prime}k^{\prime}}$ given in (\ref{eigenvaluesLikeProblemUV}).
A more general $S_{jk;j^{\prime}k^{\prime}}$ could potentially generate richer dynamics.
For this reason, we believe that the first step in the roadmap for superstate quantum computations
would involve solving the stationary problem of maximizing the fidelity (\ref{Fidelity}).
The most speculative question in this context is whether there exists a physical process that,
for a specially constructed $S_{jk;j^{\prime}k^{\prime}}$, could yield the pair $\lambda^{[0]}_{ij},\mathcal{U}^{[0]}_{jk}$
in a single measurement act -- thereby solving the quantum inverse problem at a physical level.

\subsection{\label{examples2D}Illustrative examples of 2D dynamics}
Consider a quantum system with a Hamiltonian $H$ that defines $\mathcal{U}$  in (\ref{Uquantum}).
Let the operators  $\mathfrak{U}$ and $\mathfrak{V}$ from (\ref{transfUDynamics}),
representing a single time step of (\ref{UtTimeEvolution}),
commute with $H$ and, consequently, with the initial $\mathcal{U}$.
In this ``all-commuting'' case, we obtain that:
\begin{itemize}
\item If $\mathfrak{U}=\mathfrak{V}$, then (\ref{transfUDynamics}) represents an identity transformation,
implying that
$\widetilde{\mathcal{U}}=\mathcal{U}$.

\item If $\widetilde{H}=H+A$, where the perturbation $A$ commutes with $H$ (e.g., if $A$ is merely an energy shift),
such as in the case where
$H=\scriptsize\begin{pmatrix}\epsilon_1 & 0 \\ 0 & \epsilon_2\end{pmatrix}$
and
$A=\scriptsize\begin{pmatrix}\delta\epsilon_1 & 0 \\ 0 & \delta\epsilon_2\end{pmatrix}$,
then the transformation (\ref{transfUDynamics}), with $\mathfrak{U}$ obtained from (\ref{Uquantum})
using $H=A$ and $\mathfrak{V}=\mathds{1}$, transforms the quantum system $H$ into $\widetilde{H}=H+A$.
\end{itemize}

In the general case where the ``all-commuting'' conditions are not satisfied,
the situation becomes more complicated. In this case,
it is convenient to consider the system in the canonical basis introduced above in Section \ref{canonicFormBasis}. We obtain:
\begin{itemize}
\item
If we take the initial state to be one of the solutions of (\ref{eigenvaluesLikeProblem}),
such as $\mathcal{U}^{[0]}$, which is the most common one,
and perform the analysis in the canonical basis where the operator $\mathcal{U}$ is the unit matrix (\ref{Udiag}),
then (\ref{transfUDynamics}) becomes an identity transformation for any arbitrary $\mathfrak{U}=\mathfrak{V}$,
without requiring any commutation conditions.
\item
If we consider the system in the canonical basis of the initial $\mathcal{U}$ state, any quantum system $\widetilde{H}$
can be obtained from the initial quantum system  $H$ through the transformation (\ref{transfUDynamics}),
with $\mathfrak{U}$ obtained from (\ref{Uquantum}) using $\widetilde{H}$
and $\mathfrak{V}=\mathds{1}$.
This holds because, in this basis, the initial state $\mathcal{U}$ is the unit matrix (\ref{Udiag}).
The obtained system is the desired one in terms of the match between the exponent of their Hamiltonians (\ref{Uquantum});
however, the Hamiltonians themselves may differ.

\item
A quantum system transformation from $H$ to $\widetilde{H}$
can also be performed without using a canonical basis. If $\mathcal{U}$ corresponds to $H$ and $\widetilde{\mathcal{U}}$
corresponds to $\widetilde{H}$ as in Eq. (\ref{Uquantum}),
then by selecting $\mathfrak{U}=\widetilde{\mathcal{U}}$
and $\mathfrak{V}=\mathcal{U}$, the transformation (\ref{transfUDynamics}) converts $\mathcal{U}$ to $\widetilde{\mathcal{U}}$.
Here, $\mathfrak{V}$ is used to ``remove'' the old quantum dynamics $H$,
while $\mathfrak{U}$ is used to ``replace'' it with the required dynamics corresponding to $\widetilde{H}$.
By selecting these ``horizontal'' $\mathfrak{U}$ and ``vertical'' $\mathfrak{V}$
transformations in the 2D diagram (\ref{qcExampleU}),
the quantum system $H$ is converted into $\widetilde{H}$,
in terms of the match between the exponents of their Hamiltonians (\ref{Uquantum}).

\end{itemize}

\section{\label{conclusion}Conclusion}

\begin{table*}[t]
\caption{\label{tabDistinction}
A comparison table of traditional quantum mechanics and superstate quantum mechanics.
}
\begin{ruledtabular}
\setlength\extrarowheight{1em}
\begin{tabular}{p{3.5cm}|p{5.1cm}p{6.5cm}}
& Quantum Mechanics
& Superstate Quantum Mechanics (SQM) \\\hline
The state
is described by
& Vector $\psi_j$ & Unitary operator ${\mathcal U}_{jk}$ \\
Constraints on state &
$1=\sum_j |\psi_j|^2$ &
$\delta_{ij}=\sum_{k} \mathcal{U}_{ik} \mathcal{U}^*_{jk}$ \\
The system is defined by &
Hamiltonian $H_{ij}$, Hermitian matrix. &
Hermitian tensor $S_{jk;j^{\prime}k^{\prime}}=S^*_{j^{\prime}k^{\prime};jk}$ \\
Stationary states &
Eigenproblem
$H \psi =\lambda \psi$,
where eigenvalue-scalar $\lambda$ is energy level. &
Algebraic problem
$S \mathcal{U} = \lambda \mathcal{U}$,
where eigenmatrix $\lambda_{ij}$ is a Hermitian matrix,
fidelity $\mathcal{F}=\mathrm{Tr} \lambda$. \\
Dynamic equation &
$\displaystyle i\hbar\frac{\partial \psi}{\partial t}= H \psi$ &
$\displaystyle i\hbar\frac{\partial \mathcal{U}}{\partial t}= S \mathcal{U}$
or $\displaystyle i\hbar\frac{\partial \mathcal{U}}{\partial t}= \Braket{\mathcal{U}|S|\mathcal{U}} \mathcal{U}$,
subject to future research.
\\
Possible time-evolution &
$\Ket{\psi^{(t)}}=U^{(t)}\dots U^{(\tau)} \Ket{\psi^{(0)}}$, 1D diagram (\ref{qcExample}) &
$\mathcal{U}^{(t)}=U^{(t)}\dots U^{(\tau)} \mathcal{U}^{(0)} V^{(\tau)\dagger}\dots V^{(t)\dagger}$, 2D diagram (\ref{qcExampleU}) \\
Pure and mixed states&
Vector $\Ket{\psi}$ and density matrix $\rho$. &
Unitary operator $\mathcal{U}$ and mixed unitary channel $\Upsilon$, Eq. (\ref{densMatrConvexSuperpos}). \\
Time evolution of
stationary solutions superposition
$\scriptstyle a \Ket{\psi^{[0]}}+b\Ket{\psi^{[1]}}$.
&
$\scriptstyle a \exp\left[-i\frac{t}{\hbar}\lambda^{[0]}\right]\Ket{\psi^{[0]}}+
b \exp\left[-i\frac{t}{\hbar}\lambda^{[1]}\right]\Ket{\psi^{[1]}}$,
where $\lambda^{[0,1]}$ are scalars.&
A linear superposition of eigenstates $\mathcal{U}^{[s]}$ may not describe a physically meaningful state.
\\
Where arises &
Physical processes in the universe. &
Quantum inverse problem,
machine learning,
computer modelling,
physical processes? \\
A single measurement act
in a physical process.
&
A scalar equal to an eigenvalue of $R$, with an expected value $R_{ex}=\Braket{\psi|R|\psi}$, where $R$ is a known Hermitian matrix.
The measurement act yields an eigenvalue of $R$,
and the wavefunction $\Ket{\psi}$ collapses to the corresponding eigenvector of $R$.
&
The fundamental question is whether there exists a physical process capable of solving
a quantum inverse problem in a single measurement act,
yielding an eigenmatrix as the result ---
analogous to how an eigenvalue arises from an operator measurement with wavefunction collapse in traditional quantum mechanics.
The post-measurement state collapse rules for a quantum channel $\mathcal{U}$ are an open question.
\end{tabular}
\end{ruledtabular}
\end{table*}

In this work, we develop a Superstate Quantum Mechanics (SQM) framework,
which considers states in Hilbert space subject to multiple quadratic constraints,
where both stationary and non-stationary problems are governed by the same tensor $S_{jk;j^{\prime}k^{\prime}}$.
Traditional quantum mechanics corresponds to states on a unit sphere,
representing a single quadratic constraint imposed by the wavefunction’s unit norm,
with both stationary and non-stationary dynamics governed by the Hamiltonian $H$.
When SQM represents the state as a unitary operator, the stationary problem corresponds to a quantum inverse problem ---
that is, identifying a quantum system given a sample of observed states
(which must be information-complete\cite{torlai2023quantum}; otherwise, the problem becomes degenerate).
This problem is equivalent to a new algebraic problem (\ref{eigenvaluesLikeProblem}),
for which an efficient computational algorithm was developed in our previous works;
the numerical algorithm for obtaining multiple solutions to the algebraic problem requires further research,
as discussed in Appendix \ref{SolutionsHierarchyConstraints}.

Among the immediate applications of the developed theory are stationary SQM problems related
to the reconstruction of unitary and projection operators and,
with an appropriate fidelity approximation, the reconstruction of general quantum channels.
In Appendix \ref{QuadraticFidelity}, we have presented a formulation of the general quantum channel
reconstruction problem as a QCQP. However, practical application of this formulation requires computational optimization.
In addition, see Appendix \ref{QCcomputationalModel} below, we consider a classical computational model
in which knowledge is represented in the form of a quantum channel.
Unlike conventional quantum computation, where simple gates are typically employed because
they are easier to realize physically, the mappings used in this classical computational framework need not be simple.

For non-stationary SQM, we consider the evolution of a (partially) unitary operator,
which induces an evolution of the quantum system itself,
beyond the standard description in terms of a time-dependent Hamiltonian $H(t)$.
This opens the possibility of developing a new class of quantum-computation algorithms based on higher-order dynamics.
Although no physical process is currently known to realize this type of quantum-system evolution,
the study of higher-order quantum-computation algorithms represents an interesting theoretical research direction.
For a linear map, this becomes a variant of higher-order quantum map \cite{taranto2025higher}.
An important result of this paper is the introduction of the Hermitian tensor $S_{jk;j^{\prime}k^{\prime}}$,
which is related to the quantum inverse problem (\ref{eigenvaluesLikeProblem}),
can be derived from system observations, and is well-suited for numerical simulations.
In contrast to regular 1D quantum computations (\ref{qcExample}), which transform the state $\Ket{\psi^{(0)}}$,
2D quantum computations (\ref{qcExampleU}) transform the quantum system itself,
allowing for the development of a variety of intriguing 2D algorithms.
This approach naturally bridges direct and inverse quantum mechanics problems.
A nonlinear map of GPE type is also considered as a candidate for the dynamic equation.
Since the unitary mapping corresponds to a quantum system with the Hamiltonian given by (\ref{logUCalc}),
the non-stationary problem of SQM dynamics ---
whether described by higher-order quantum map or a GPE type equation ---
can be interpreted as the evolution of the quantum system itself.
The comparison of traditional quantum mechanics and Superstate Quantum Mechanics is presented in Table \ref{tabDistinction}.

\begin{acknowledgments}
This research was supported by Autretech Group,
a resident company of the Skolkovo Technopark.
  We thank our colleagues from the Autretech R\&D department
  who provided insight and expertise that greatly assisted the research.
  Our grateful thanks are also extended
  to Mr. Gennady Belov for his methodological support in doing the data analysis.

In this work, we present an analytical study of a new form of quantum mechanics,
an approach that has arisen from the numerical study of quantum inverse problems.
Iya Pavlovna Ipatova, the first scientific advisor of A.M., O.P., and V.M.,
always emphasized the synergy of both analytical and numerical approaches.
For many years, she worked at the
\href{https://www.ioffe.ru/}{Ioffe Institute}
and taught physics (together with V.D. and V.I.) at
\href{https://english.spbstu.ru/}{St.Petersburg Polytechnic University}.
The authors recognize her ideological contribution to this work.
The article was written based on the presentation at the
colloquium
on the occasion of the 95th anniversary
of Iya Pavlovna Ipatova's\cite{birman2004ija} birthday, held on December 19, 2024.
This work is dedicated to her memory.
\end{acknowledgments}

\appendix
\section{\label{SolutionsHierarchyConstraints}Numerical Construction of a Hierarchy of Stationary Solutions}

In a regular eigenproblem (\ref{stationarySchrodinger})
the eigenvectors can be obtained by solving a hierarchy of optimization
problems subject to the linear constraints (\ref{constraintsDensMatrixMaxH}).
\begin{align}
\lambda^{[s]}&=\frac{
\Braket{\phi|H|\phi}
}{
\Braket{\phi|\phi}
}\xrightarrow[{\mathcal{\phi}}]{\quad }\max \label{DensMatrixMaxH} \\
0&=\Braket{\phi|\phi^{[s^{\prime}]}} & s^{\prime}<s \label{constraintsDensMatrixMaxH}
\end{align}

The construction of a hierarchy of solutions for the algebraic problem (\ref{eigenvaluesLikeProblem})
is more difficult. There is no issue for the ground state solution $\mathcal{U}^{[0]}$,
as this corresponds to the optimization problem (\ref{Fidelity})
without the condition of a previous state solution, 
these are referred to as external linear constraints in \cite{belov2024quantumPRE}.
Whereas in the eigenproblem the hierarchy linear constraints (\ref{constraintsDensMatrixMaxH})
are the same for both
$\Braket{\phi|\phi^{[s^{\prime}]}}$ and $\Braket{\phi|H|\phi^{[s^{\prime}]}}$,
in the algebraic problem (\ref{eigenvaluesLikeProblem})
the corresponding linear constraints differ.
A few examples of the constraints that can be used to construct the solution hierarchy.
\begin{align}
0&=\Braket{\mathcal{U}^{[s^{\prime}]}|\lambda^{[s^{\prime}]}|\mathcal{U}}
\label{constraintsUPreviousNumerator} \\
0&=\Braket{\mathcal{U}^{[s^{\prime}]}|\mathcal{U}}
\label{constraintsUPreviousDeniminator} \\
0&=\Braket{\mathcal{U}^{[s^{\prime}]}|S|\mathcal{U}}
\label{constraintsUPreviousOrthogonalChannel}
\end{align}
The solutions created with (\ref{constraintsUPreviousNumerator}),
(\ref{constraintsUPreviousDeniminator}), or (\ref{constraintsUPreviousOrthogonalChannel})
are all different because $\lambda$ is now a Hermitian matrix.
As discussed in \cite{belov2024quantumPRE},
the approach (\ref{constraintsUPreviousOrthogonalChannel}),
which corresponds to states orthogonality condition (\ref{otrConditionS0}),
allows for the separation of contributions to fidelity (\ref{qcAsSum}).
However, the difficulty arises from the fact that, for $s > 0$,
Eq. (\ref{eigenvaluesLikeProblemSlev}) is not satisfied exactly.
Instead, it only implies that the projections onto the $\mathcal{U}$-space,
which satisfies the external constraints, must match.
For example, even if we impose the constraint
$0 = \Braket{\mathcal{U}^{[0]}|S|\mathcal{U}^{[1]}}$ (\ref{constraintsUPreviousOrthogonalChannel}),
this does not necessarily lead to
$0 = \Braket{\mathcal{U}^{[0]}|\lambda^{[1]}|\mathcal{U}^{[1]}}$
because $S \mathcal{U}^{[0]} = \lambda^{[0]} \mathcal{U}^{[0]}$.

This creates a difficulty in studying dynamics, where it is convenient for Eq. (\ref{eigenvaluesLikeProblemSlev})
to hold exactly for all $s$.
As shown above, to satisfy Eq. (\ref{eigenvaluesLikeProblemSlev}) exactly,
the linear constraint must take the form (\ref{otrCondition}). Explicitly, in matrix notation, this is
\begin{align}
\sum\limits_{j,j^{\prime}=0}^{D-1}\sum\limits_{k=0}^{n-1}
\left(\lambda_{jj^{\prime}}^{[s^{\prime}]} \mathcal{U}^{[s^{\prime}]}_{j^{\prime}k}\right)^* \mathcal{U}_{jk}
&=
\sum\limits_{j,j^{\prime}=0}^{D-1}\sum\limits_{k=0}^{n-1}
\mathcal{U}^{[s^{\prime}]*}_{jk}
\lambda_{jj^{\prime}}^{[s]}\mathcal{U}_{j^{\prime}k}
\label{otrConditionExplicitMatrixS}
\end{align}
for $s^{\prime}<s$, where we seek the $s$-th solution.
The difficulty with Eq. (\ref{otrConditionExplicitMatrixS}) is that the value of $\lambda^{[s]}$,
which is required for the constraint, is not yet known.
While the constraint (\ref{constraintsUPreviousOrthogonalChannel}) considered in \cite{belov2024quantumPRE}
is a linear function of $\mathcal{U}_{jk}$ with precisely known coefficients,
the constraint (\ref{otrConditionExplicitMatrixS}) has coefficients by $\mathcal{U}_{jk}$
that depend on the yet unknown $\lambda^{[s]}$, which itself is calculated from $\mathcal{U}_{jk}$.
A trivial solution is to use the current iteration $\lambda$ as $\lambda^{[s]}$  in the constraints,
but this method does not always converge.
The major issue with it is that, assuming $\lambda^{[s]}=\lambda$, the already obtained solutions $s^{\prime}=0\dots s-1$
may again contribute to the sought solution $s$. This creates an issue in numerical implementation.
However, to demonstrate the concept, the code from \cite{belov2024quantumPRE} was modified to support
the external constraints (\ref{otrConditionExplicitMatrixS}), along with other types; see the class \texttt{\seqsplit{com/polytechnik/kgo/PreviousSolutionOrthogonalityConditions.java}}. The demo usage example,
\texttt{\seqsplit{com/polytechnik/algorithms/DemoDMGeneralMappingTest.java}},
now runs with all possible types of constraints implemented.
All self-tests are disabled when the constraints (\ref{otrConditionExplicitMatrixS}) are used,
as convergence may be poor.
Although the current numerical implementation for constructing the hierarchy (\ref{otrConditionExplicitMatrixS})
converges only for certain $S_{jk;j^{\prime}k^{\prime}}$,
it demonstrates the validity of the approach that utilizes the $\lambda$-dependent orthogonality condition
(\ref{otrCondition}) to construct a hierarchy of solutions.
For example, see a test run:
\texttt{\seqsplit{java\ com/polytechnik/algorithms/DemoDMGeneralMappingTest\ 2>\&1\ |\ grep\ fdiffSK}},
where a hierarchy of three solutions is constructed.
For orthogonality (\ref{otrConditionExplicitMatrixS}), which is denoted as
\texttt{\seqsplit{orthogonalityType=L1U1U0mU1L0U0}},
the hierarchies of three solutions for the four different $S_{jk;j^{\prime}k^{\prime}}$ are:
\begin{verbatim}
orthogonalityType=L1U1U0mU1L0U0 solution#=0
     fdiffSK=3.766333749746028E-24
     Tr lambda0=410.59157706919996 flagOK=true
orthogonalityType=L1U1U0mU1L0U0 solution#=1
     fdiffSK=1.9896185359584374E-9
     Tr lambda1=317.3980075465856 flagOK=true
orthogonalityType=L1U1U0mU1L0U0 solution#=2
     fdiffSK=2.1065565301784156E-8
     Tr lambda2=252.44684491981542 flagOK=true

orthogonalityType=L1U1U0mU1L0U0 solution#=0
     fdiffSK=7.65025644955303E-25
     Tr lambda0=652.4722048564872 flagOK=true
orthogonalityType=L1U1U0mU1L0U0 solution#=1
     fdiffSK=6.325612650989948E-7
     Tr lambda1=506.8243802433262 flagOK=true
orthogonalityType=L1U1U0mU1L0U0 solution#=2
     fdiffSK=7.907952725589345E-7
     Tr lambda2=403.8568961643975 flagOK=true

orthogonalityType=L1U1U0mU1L0U0 solution#=0
     fdiffSK=3.798582634500135E-25
     Tr lambda0=1039.4442877993176 flagOK=true
orthogonalityType=L1U1U0mU1L0U0 solution#=1
     fdiffSK=71.88802400224358
     Tr lambda1=811.5827439382526 flagOK=false
orthogonalityType=L1U1U0mU1L0U0 solution#=2
     fdiffSK=172.53103944532708
     Tr lambda2=656.7028629629301 flagOK=false

orthogonalityType=L1U1U0mU1L0U0 solution#=0
     fdiffSK=2.510130039767752E-25
     Tr lambda0=627.044255728697 flagOK=true
orthogonalityType=L1U1U0mU1L0U0 solution#=1
     fdiffSK=236.92078517799965
     Tr lambda1=503.1963951381101 flagOK=false
orthogonalityType=L1U1U0mU1L0U0 solution#=2
     fdiffSK=212.59370321253004
     Tr lambda2=500.557762368503 flagOK=false
\end{verbatim}
Here, \verb+fdiffSK+ represents the $L^2$ norm of the vector $S \mathcal{U}-\lambda \mathcal{U}$.
For the first two $S_{jk;j^{\prime}k^{\prime}}$ the algorithm converges and provides good accuracy,
but it fails to converge for the last two. This issue arises from the more complex form of the constraint in Eq. (\ref{otrConditionExplicitMatrixS})
and requires further work on the numerical implementation.
Nonetheless, test runs clearly demonstrate the correctness of the orthogonality of solutions in the form
of Eq. (\ref{otrCondition}) and the ability to construct a hierarchy of solutions
for the problem in Eq. (\ref{eigenvaluesLikeProblem})
with the constraints (\ref{otrConditionExplicitMatrixS}).

When it is acceptable for a solution $\mathcal{U}^{[s]}_{jk}$  to have Eq. (\ref{eigenvaluesLikeProblemSlev})
that does not satisfy some projections, the hierarchy (\ref{constraintsUPreviousOrthogonalChannel}),
corresponding to the orthogonality condition (\ref{otrConditionS0}),
is much preferred for reasons of numerical implementation and the possibility of expansion in (\ref{expansionSinUs}).
A similar run for orthogonality (\ref{constraintsUPreviousOrthogonalChannel}),
which is denoted as
\texttt{\seqsplit{orthogonalityType=USU}},
successfully finds the hierarchies of three solutions for the four different $S_{jk;j^{\prime}k^{\prime}}$.
For $D=n=10$, no problems with numerical stability arise when increasing the number of solutions in the hierarchy up to seven.
For further increases in $N_s$, the problem of resolving conflicts between
the external constraints (\ref{constraintsUPreviousOrthogonalChannel})
and the adjustment of $\mathcal{U}^{[s]}_{jk}$ iteration to unitarity requires
additional attention in the numerical implementation \cite{belov2024quantumPRE}.

There is another way to create multiple solutions without using explicit orthogonality conditions.
In our numerical algorithm \cite{belov2024partiallyPRE},
in the eigenstate selection step,
selecting the largest $\mu^{[s]}$, second largest $\mu^{[s]}$, third largest $\mu^{[s]}$, etc.,
converges to a different solution of the original problem.
This way, we managed to obtain up to a dozen different solutions.
Starting with about the fifth largest eigenvalue, convergence may not always be observed.
This is an alternative method to obtain several solutions of (\ref{eigenvaluesLikeProblem}).

To study various aspects of quantum channel dynamics,
we require multiple solutions of Eq. (\ref{eigenvaluesLikeProblem})
that resemble the stationary solutions of the Schr\"{o}dinger equation.
Currently, we have two distinct methods outlined above for obtaining these stationary solutions.

Note that the currently available software \cite{polynomialcode}
is implemented only in real space $\mathcal{U}$ (orthogonal matrix).
The complex space $\mathcal{U}$ (unitary matrix) can be directly obtained by doubling the problem's
dimension and considering the combined real vector
$(\mathrm{Re}\, \mathcal{U}, \mathrm{Im}\, \mathcal{U})$.
This real problem of dimension $2Dn$ has the same quadratic fidelity (\ref{Fidelity})
and quadratic constraints (\ref{unitarityCond}) and can be solved using the same method
from Ref. \cite{belov2024partiallyPRE}, which employs Lagrange multipliers with convergence-helper linear constraints.
This approach creates $2D(D-1)/2$ equations for off-diagonal elements set to zero (real and imaginary parts)
and $2D$ equations for diagonal elements set to one
(actually $D$ since the diagonal elements in (\ref{GramMatrix}) are all real).
The $D(D-1)+2D$ quadratic conditions of unitarity are then converted to a simplified (partial)
constraint (\ref{unitarityCondSimplified}) and $D(D-1)+2D-2$ convergence-helper homogeneous linear constraints.
While considering a real-valued optimization problem of dimension $2Dn$
may not be the most efficient approach from a computational complexity standpoint,
it directly enables obtaining the numerical solution in complex space using only real-space software.
If complex arithmetic is available, the optimization problem can be formulated as a
complex-valued problem of dimension $Dn$,
with $D(D-1)/2+D-1$ complex-valued convergence-helper homogeneous linear constraints.
This formulation involves an iterative step that requires solving a Hermitian matrix eigenproblem
of dimension $Dn-(D(D-1)/2+D-1)$,
which is the same as for the real-valued problem with orthogonal $\mathcal{U}$.

\section{\label{dualProblem}Dual Optimization Problem}
For an optimization problem with a Lagrangian, such as $\mathcal{L}$ (\ref{lagrangetovariateNUDlen}),
the solution $\lambda,\mathcal{U}$ corresponding to the original constrained optimization
is always a saddle point of the Lagrangian function\cite{brezzi1974existence}.
The problem of maximizing $\mathcal{L}$ over $\mathcal{U}$ is known as the primal optimization problem.
There always exists a dual problem\cite{bazaraa2006nonlinear}, corresponding to the minimization of $\mathcal{L}$ over $\lambda$.
When a problem has a ``zero duality gap'', as in linear programming,
it means that the optimal values of the primal and dual problems are equal.
This indicates strong duality, where the dual problem provides a tight lower bound for the primal problem.
In the case of QCQP, the situation is not as straightforward as it is in linear programming
because dealing with the constraints of the dual problem is more complex.
The degeneracy of the Hessian matrix in the combined space of $\lambda$ and $\mathcal{U}$
makes a direct approach to the dual problem impractical for applications.
However, we can formulate a problem similar to the dual problem that allows us to calculate an approximation of the Lagrange multipliers. Varying $\mathcal{L}$ (\ref{lagrangetovariateNUDlen}) over $\mathcal{U}_{jk}$ yields $Dn$ equations,
which correspond to Eq. (\ref{eigenvaluesLikeProblem}).
\begin{align}
\frac{\delta \mathcal{L}}{\delta \mathcal{U}_{jk}}&=0
\label{varLeq0}
\end{align}
When the $\lambda,\mathcal{U}$ is
a solution to (\ref{eigenvaluesLikeProblem}), all $Dn$ equations are satisfied.
If $\mathcal{U}$ satisfies (\ref{unitarityCond}) but is not an extremum of (\ref{Fidelity}),
$\lambda$, with its $D(D+1)/2$ independent components, cannot be chosen to satisfy all $Dn$ equations.
Now, consider a dual-like problem.
\begin{align}
\sum\limits_{j=0}^{D-1}\sum\limits_{k=0}^{n-1}\left|\frac{\delta \mathcal{L}}{\delta \mathcal{U}_{jk}} \right|^2 &\xrightarrow[\lambda]{\quad }\min
\label{varLeqVariationMin}
\end{align}
It is a quadratic function of $\lambda$, and its minimization is reduced to solving a linear system. This approach can be generalized to various sets of constraints and multiple forms of Lagrange multipliers\cite{belov2024quantumPRE}.
In the case of the constraints given by (\ref{unitarityCond}),
we obtain the familiar expression for $\lambda_{ji}$ in (\ref{lambdaFromU}).
Its meaning is equivalent to estimating $\lambda = \Braket{\psi|H|\psi}$ in a regular eigenvalue problem for a given iteration value $\Ket{\psi}$.

The dual Lagrangian can be expressed in the form
\begin{align}
\widetilde{\mathcal{L}}&=
\sum\limits_{j,j^{\prime}=0}^{D-1}\sum\limits_{k,k^{\prime}=0}^{n-1}
\left(\frac{\delta \mathcal{L}}{\delta \mathcal{U}_{jk}}\right)^*
Q_{jk;j^{\prime}k^{\prime}}
\frac{\delta \mathcal{L}}{\delta \mathcal{U}_{j^{\prime}k^{\prime}}}
\label{varLeqVariationMinDual}
\end{align}
where $Q_{jk;j^{\prime}k^{\prime}}$ is an arbitrary positively definite Hermitian tensor.
For example, it may be taken as a function of the inverse of $S_{jk;j^{\prime}k^{\prime}}$,
or as $Q_{jk;j^{\prime}k^{\prime}}=\delta_{jj^{\prime}}\delta_{kk^{\prime}}$ in (\ref{varLeqVariationMin}).
The constraint is that $\lambda$ must be a Hermitian matrix.
The problem involving $\widetilde{\mathcal{L}}$ is not a true dual problem to $\mathcal{L}$ (\ref{lagrangetovariateNUDlen}),
but it has similar properties: we need to find the minimum of $\widetilde{\mathcal{L}}$ with respect to $\lambda$,
and at the extremum
$\lambda^{[s]},\mathcal{U}^{[s]}$,
the value of $\widetilde{\mathcal{L}}$ is zero.
An important advantage is that this ``dual'' problem is easy to solve --
it can be reduced to a linear system -- and can be directly incorporated into numerical methods.

\section{\label{QCcomputationalModel}Quantum Channel as Classical Computational Model}

In quantum computing, the computational model is based on the evolution of quantum systems.
It typically involves constructing a series of simple gates, connecting them in a way similar to  circuit (\ref{qcExample}).
Then, for a given initial state, the quantum system evolution produces the desired computational result.
The gates are typically chosen to be simple, as it is assumed that simpler gates are easier to implement in a physical system.
An important question is whether these gates form a ``full basis'',
meaning they can approximate any arbitrary quantum gate to any desired precision.

In this appendix, we introduce a quantum-style approach to classical computation,
in which the transformation (\ref{KrausOperator}) is performed on a classical computer,
possibly with hardware optimization.
We examine neural network techniques and, based on this analysis,
outline the specific requirements for performing classical computations that use a quantum channel
as knowledge representation.
These requirements will differ significantly from those of quantum computers:
\begin{itemize}
\item The ``gates'' are not required to be simple.
Similar to how perceptron weights, combined with an activation function, create a specialized component,
we can ``program'' each component in a highly customized manner.
\item
The model should be capable of reducing the problem’s dimensionality.
Similar to deep learning, where each layer captures specific features of the input data,
the model should also reduce the problem’s dimension.
This suggests that unitary evolution (\ref{qcExample}) is likely not the optimal choice.
The best alternative, we expect, is a trace-preserving general quantum channel (\ref{KrausOperator}).
However, as noted in Section \ref{FidelityFormulation} above, formulating the fidelity as a quadratic form can be problematic.
In this case, a trace-decreasing, partially unitary ($D<n$) map (\ref{operatorTransformU})
could potentially be considered instead.\footnote{
While simplifying computations, the partially unitary model may introduce artifacts.
See Section ``VI. A Demonstration of Partially Unitary Behavior ($D < n$)'' of Ref. \cite{belov2024partiallyPRE}.
The main advantage of the Rayleigh-quotient fidelity (\ref{fidelityProjectionsApproximationBBSProjectionsU})
is that it avoids artifacts and enables the exact reconstruction of projection operators \cite{belov2026semidefinitePRE}.
}
We believe the computational model should take a density matrix as input and produce a density matrix as output, making a general quantum channel the best candidate for this task.
For classical computations, the fidelity between a mixed state and a pure state $\Braket{\phi|\varrho|\phi}$
is the most promising choice, since it captures most of the important effects in ML and AI,
while at the same time being a quadratic form in the mapping operators.
\item
The concepts of an ensemble of predictors, voting, and feature extraction involve combining entities
of different types into a single result.
When these inputs are represented as density matrices of different types,
the combined result can be represented as a density matrix in the space corresponding
to the direct product of the input spaces of all input density matrices.
\item
For any computational model, the total computation should be separable into a number of simpler steps,
creating a hierarchy of complexity. The algorithm should be divisible into simpler subroutines,
which, in turn, can be split into even simpler ones, and so on, across multiple levels.
Based on typical software nesting levels of 10-40 and the 10-100 layers commonly found in deep learning models,
we estimate that the required number of levels in the hierarchy should be at least 10.
\end{itemize}
We should emphasize that there are currently two clearly distinct approaches to knowledge representation based on unitary operators:
\begin{itemize}
\item
Quantum computation, in which an algorithm consists of a sequence of unitary transformations applied to an initial quantum state.
The dimension of the equivalent Hilbert space typically grows exponentially with the number of qubits, namely as $2^{n_q}$
for $n_q$ qubits.
\item
Unitary learning in machine learning and artificial intelligence \cite{bisio2010optimal,arjovsky2016unitary,hyland2017learning,huang2021learning,yu2023optimal}, in which input features are encoded into an initial state,
transformed by a unitary operator, and subsequently decoded to extract the desired features.
This approach is typically implemented on classical computers, and the problem dimension is usually below several hundred thousand.
\end{itemize}
The present framework is related to the second category,
in which unitary operators are generalized to quantum channels that serve as representations
of knowledge and are reconstructed from data through a quantum inverse problem.
Since we are considering classical computers, it is more convenient to work in \hyperref[vectorizedSpaceFN]{vector space}
rather than qubit space.
The input vector of dimension $n$ effectively corresponds to $\log_2 n$ input qubits.
The computational model consists in creating a complex quantum channel that takes a density matrix of dimension $n$
as input and outputs a density matrix of smaller dimension $D$.
This output is represented as a combination of simple quantum channels arranged in a hierarchical manner.

The first required component is a quantum channel $\mathcal{Q}$
that takes a density matrix of dimension $n_{\mathcal{Q}}$
as input and outputs a density matrix of dimension
$D_{\mathcal{Q}}$.
Based on the requirements above, for classical computation, it is convenient to have
$n_{\mathcal{Q}}\gg D_{\mathcal{Q}}$.
An example of such a channel $\mathcal{Q}$
could be an extractor of specific features from the input data.
\begin{equation}
\hbox{
\begin{circuitikz}[line width=1pt]
\tikzstyle{operator} = [draw,fill=white,minimum size=1.5em] 
\tikzstyle{phase} = [draw,fill,shape=circle,minimum size=5pt,inner sep=0pt]
\tikzstyle{Xcross} = [path picture={ 
\draw[thick,black,inner sep=0pt]
(path picture bounding box.south east) -- (path picture bounding box.north west) (path picture bounding box.south west) -- (path picture bounding box.north east);
}]
\tikzstyle{cross} = [path picture={ 
\draw[thick,black](path picture bounding box.north) -- (path picture bounding box.south) (path picture bounding box.west) -- (path picture bounding box.east);
}]
\tikzstyle{Ocross} = [draw,circle,cross,minimum width=0.3 cm]

\ctikzset{multipoles/thickness=2}
\ctikzset{multipoles/external pins thickness=2}
\ctikzset{multipoles/qfpchip/pin spacing=0.6}

\node [qfpchip,
line width=0.6pt,
num pins=4, no topmark,
external pins width=0.1,
hide numbers,
draw only pins={1,3}
](C){{\Huge $\mathcal{Q}$}};

\node [left, font=\Large] at (C.pin 1) {$\rho$};

\node [right, font=\Large] at (C.pin 3) {$\varrho$};

\end{circuitikz}
}
\label{qcClassicQ}
\end{equation}
This component converts a density matrix $\rho$ of dimension $n_{\mathcal{Q}}$
to a density matrix  $\varrho$ of dimension $D_{\mathcal{Q}}$ according to the quantum channel mapping (\ref{KrausOperator})
(or possibly (\ref{operatorTransformU}) for a partially unitary mapping $D<n$).

The second required component is the direct (tensor) product multiplier $\mathcal{T}$.
\begin{equation}
\hbox{
\begin{circuitikz}[line width=1pt]
\tikzstyle{operator} = [draw,fill=white,minimum size=1.5em] 
\tikzstyle{phase} = [draw,fill,shape=circle,minimum size=5pt,inner sep=0pt]
\tikzstyle{Xcross} = [path picture={ 
\draw[thick,black,inner sep=0pt]
(path picture bounding box.south east) -- (path picture bounding box.north west) (path picture bounding box.south west) -- (path picture bounding box.north east);
}]
\tikzstyle{cross} = [path picture={ 
\draw[thick,black](path picture bounding box.north) -- (path picture bounding box.south) (path picture bounding box.west) -- (path picture bounding box.east);
}]
\tikzstyle{Ocross} = [draw,circle,cross,minimum width=0.3 cm]

\ctikzset{multipoles/thickness=2}
\ctikzset{multipoles/dipchip/width=0.7}
\ctikzset{multipoles/external pins thickness=2}
\ctikzset{multipoles/dipchip/pin spacing=0.7}
\draw (0,0) node[dipchip,
line width=0.6pt,
num pins=6, no topmark,
external pins width=0.1,
hide numbers
,draw only pins={1,3,5,}
](C){{\Huge $\mathcal{T}$}};

\node [left, font=\Large] at (C.pin 1) {$\rho^{(1)}$};
\node [left, font=\Large] at (C.pin 2) {\dots};
\node [left, font=\Large] at (C.pin 3) {$\rho^{(m)}$};

\node [right, font=\Large] at (C.pin 5) {$\varrho$};

\end{circuitikz}
}
\label{qcClassicT}
\end{equation}
It takes $m$ input density matrices $\rho^{(s)}$, $s=1\dots m$,
and creates an output density matrix $\varrho$ with a dimension equal to the product of the input matrices' dimensions,
$\prod\limits_{s=1}^{m} n_{\rho^{(l)}}$, corresponding to the multi-index $\mathbf{j}=(j_1,\dots,j_m)$.
\begin{align}
\varrho_{\mathbf{j}\mathbf{j}^{\prime}}&=
\rho^{(1)}_{j_1j_1^{\prime}} \dots \rho^{(m)}_{j_mj_m^{\prime}}
\label{prodRho}
\end{align}
If the basis of the input sample is not orthogonal,
this expression would involve some orthogonalization procedure,
as discussed in Appendix A and E of Ref. \cite{malyshkin2019radonnikodym}.
In all calculations below, we will assume that the orthogonalization procedure has already been applied to the initial basis,
and the joint density matrix of several input attributes is given by (\ref{prodRho}).
Since the invariance property of the theory, the specific choice of the orthogonalization procedure
is only a question of numerical stability.

The idea is to represent a general $A^{IN}/A^{OUT}$ quantum channel (\ref{KrausOperator}),
which can be graphically represented as (\ref{qcClassicQ}), as a hierarchical combination of simpler quantum channels,
similar to the structure of a neural network,
with the goal of creating a system that can be efficiently simulated on a classical computer.
A simple tree-layer system is presented in the circuit:
\begin{equation}
\hbox{
\begin{circuitikz}[line width=1pt]
\tikzstyle{operator} = [draw,fill=white,minimum size=1.5em] 
\tikzstyle{phase} = [draw,fill,shape=circle,minimum size=5pt,inner sep=0pt]
\tikzstyle{Xcross} = [path picture={ 
\draw[thick,black,inner sep=0pt]
(path picture bounding box.south east) -- (path picture bounding box.north west) (path picture bounding box.south west) -- (path picture bounding box.north east);
}]
\tikzstyle{cross} = [path picture={ 
\draw[thick,black](path picture bounding box.north) -- (path picture bounding box.south) (path picture bounding box.west) -- (path picture bounding box.east);
}]
\tikzstyle{Ocross} = [draw,circle,cross,minimum width=0.3 cm]

\ctikzset{multipoles/thickness=2}
\ctikzset{multipoles/dipchip/width=0.4}
\ctikzset{multipoles/external pins thickness=2}
\ctikzset{multipoles/dipchip/pin spacing=0.6}
\draw (0,0) node[dipchip,
line width=0.6pt,
num pins=10, no topmark,
external pins width=0.1,
hide numbers
,draw only pins={1,2,3,4,5,8}
](C){{\Large $\mathcal{T}$}};

\draw (C.pin 1) to[short,-] ++(-0.6,0) node[operator] (H1) {\Large $\mathcal{Q}_{11}$} ;
\draw (C.pin 2) to[short,-] ++(-0.6,0) node[operator] (H2) {\Large $\mathcal{Q}_{12}$} ;
\draw (C.pin 3) to[short,-] ++(-0.6,0) node[operator] (H3) {\Large $\mathcal{Q}_{13}$} ;
\draw (C.pin 4) to[short,-] ++(-0.6,0) node[operator] (H4) {\Large $\mathcal{Q}_{14}$} ;
\draw (C.pin 5) to[short,-] ++(-0.6,0) node[operator] (H5) {\Large $\mathcal{Q}_{15}$} ;

\draw (H1.west) to[short,-] ++(-0.2,0) |- (H5);
\draw (H2.west) to[short,-] ++(-0.2,0);
\draw (H4.west) to[short,-] ++(-0.2,0);
\draw (H5.west) to[short,-] ++(-0.2,0);
\draw (H3.west) to[short,-*] ++(-0.4,0) to node[anchor=south east]{$IN$} ++(0.4,0) ;

\draw (C.pin 8) ++(2.3,0) node[dipchip,
line width=0.6pt,
num pins=10, no topmark,
external pins width=0.1,
hide numbers
,draw only pins={1,2,3,4,5,8}
](CC){{\Large $\mathcal{T}$}};
\draw (CC.pin 3) to[short,-] ++(-0.6,0) node[operator] (HH3) {\Large $\mathcal{Q}_{21}$} to[short,-] (C.pin 8) ;
\draw (CC.pin 1) to[short,-] ++(-0.6,0) node[operator] (HH1) {\Large $\mathcal{Q}_{22}$}  ;
\draw (CC.pin 2) to[short,-] ++(-0.6,0) node[operator] (HH2) {\Large $\mathcal{Q}_{23}$} ;
\draw (CC.pin 4) to[short,-] ++(-0.6,0) node[operator] (HH4) {\Large $\mathcal{Q}_{24}$}  ;
\draw (CC.pin 5) to[short,-] ++(-0.6,0) node[operator] (HH5) {\Large $\mathcal{Q}_{25}$}  ;
\draw (HH1.west) to[short,-] ++(-0.2,0) |- (HH5);
\draw (HH2.west) to[short,-] ++(-0.2,0);
\draw (HH4.west) to[short,-] ++(-0.2,0);
\draw (HH5.west) to[short,-] ++(-0.2,0);
\draw (CC.pin 8) -- ++(0.6,0) node[operator] (HHH3) {\Large $\mathcal{Q}_{31}$}  to node[anchor=south west]{$OUT$} ++(0.8,0) to[short,-*] ++(0,0);

\end{circuitikz}
}
\label{qcExampleQCClassic}
\end{equation}
Here, the $\mathcal{Q}_{rs}$
represent simpler quantum channels combined in layers, the
$s$-th quantum channel in the 
$r$-th layer.
Each such operator may either reduce or leave unchanged the dimension of the density matrix it transforms.
The operator $\mathcal{T}$ 
is given by (\ref{prodRho}) and combines the spaces corresponding to different density matrices.
This transformation may potentially cause degeneracy issues.
The form shown in (\ref{qcExampleQCClassic}) represents a general quantum channel that is convenient for modeling on classical computers.
For example, some of the $\mathcal{Q}_{rs}$ 
components can be created or learned separately from different input datasets.
A trivial version of this topology involves reducing the problem space in several steps,
such as when an input density matrix of large dimension is converted into an output density matrix
of much smaller dimension by consecutively applying three quantum channels.
\begin{equation}
\hbox{
\begin{circuitikz}[line width=1pt]
\tikzstyle{operator} = [draw,fill=white,minimum size=1.5em] 
\tikzstyle{phase} = [draw,fill,shape=circle,minimum size=5pt,inner sep=0pt]
\tikzstyle{Xcross} = [path picture={ 
\draw[thick,black,inner sep=0pt]
(path picture bounding box.south east) -- (path picture bounding box.north west) (path picture bounding box.south west) -- (path picture bounding box.north east);
}]
\tikzstyle{cross} = [path picture={ 
\draw[thick,black](path picture bounding box.north) -- (path picture bounding box.south) (path picture bounding box.west) -- (path picture bounding box.east);
}]
\tikzstyle{Ocross} = [draw,circle,cross,minimum width=0.3 cm]

\ctikzset{multipoles/thickness=2}
\ctikzset{multipoles/dipchip/width=0.4}
\ctikzset{multipoles/external pins thickness=2}
\ctikzset{multipoles/dipchip/pin spacing=0.6}
\draw  node[anchor=south east]{$IN$} (0,0) to[short,*-] ++(0.8,0) node[operator] (H1) {\Large $\mathcal{Q}_{1}$}
 -- ++(1.4,0) node[operator] (H2) {\Large $\mathcal{Q}_{2}$}
 -- ++(1.4,0) node[operator] (H3) {\Large $\mathcal{Q}_{3}$}
 to[short,-*] ++(0.8,0) to node[anchor=south west]{$OUT$} ++(0,0);

\end{circuitikz}
}
\label{qcExampleQCClassicLine}
\end{equation}
In this appendix, the lines in the circuits correspond not to a qubit,
but to a density matrix of some dimension.
The entire circuit can be considered as a flow of density matrix transformations,
and the entire framework can be referred to as a density matrix network.
The unitary learning \cite{arjovsky2016unitary} can be considered a predecessor of the density matrix network.
We see the following advantages of the proposed approach

\begin{itemize}
\item
There is a clear meaning behind the transformation -- namely, a quantum channel.
Therefore, the specific topology chosen is simply a matter of computational optimization.
For smaller dimensions, the problem can always be optimized directly without assuming any particular topology.
This contrasts with neural networks, where selecting the proper topology is a crucial factor for success.
Another reason for choosing a quantum channel is that practical data analysis typically involves both coherent and non-coherent effects,
which can be naturally captured by applying a quantum channel of the form (\ref{KrausOperator}) to classical data transformations.
We believe that representing knowledge in the form of a quantum channel allows one to model a wide variety of data types.
\item
In our opinion, the major disadvantage of neural networks is that the invariance group is typically not known.
If we transform input attributes as $\mathbf{x}^{\prime}=T\mathbf{x}$,
where $T$ is a non-degenerate linear transform,
and perform the training on the transformed attributes, neural networks typically produce a different solution than the one on the original sample.
In the density matrix network, the input/output attributes $\mathbf{x}\to\mathbf{f}$ mapping,
a sample of $M$ observations
\begin{align}
\mathbf{x}^{(l)}&\to\mathbf{f}^{(l)} & l=1\dots M
\label{xfmap}
\end{align}
is converted to the corresponding $IN/OUT$ wavefunction mapping (\ref{purePsiMap}),
for all $l=1\dots M$ \cite{malyshkin2019radonnikodym, belov2024partiallyPRE}.
\begin{align}
 \frac{\sum\limits_{j,k=0}^{n-1}x_jG^{\mathbf{x};\,-1}_{jk}x^{(l)}_k}
           {\sqrt{\sum\limits_{j,k=0}^{n-1}x^{(l)}_jG^{\mathbf{x};\,-1}_{jk}x^{(l)}_k}}
  &\to
 \frac{\sum\limits_{j,k=0}^{D-1}f_jG^{\mathbf{f};\,-1}_{jk}f^{(l)}_k}
           {\sqrt{\sum\limits_{j,k=0}^{D-1}f^{(l)}_jG^{\mathbf{f};\,-1}_{jk}f^{(l)}_k}}
  \label{psiXFGflocalizedAppendix}
\end{align}
This type of transformation is an important preliminary step for converting
a general vector-to-vector mapping (\ref{xfmap}) into entities that have the meaning
of wavefunctions (or density matrices) that can be passed through a quantum channel.
Note that, in most existing work on unitary learning,
authors simply normalize input and output vectors to unit norm and treat them as a form of
``wavefunction''; in particular, such a ``wavefunction'' does not provide proper gauge invariance.
In (\ref{psiXFGflocalizedAppendix}), the $x^{(l)}_k$ and $f^{(l)}_j$ 
are the components of vectors in a given sample (\ref{xfmap}), and
$x_k$ and $f_j$  
are the components of the arguments of the wavefunctions
$\psi^{(l)}(\mathbf{x})\to\phi^{(l)}(\mathbf{f})$ (\ref{purePsiMap}).
The matrices
$G^{\mathbf{x}}_{jk}=\Braket{x_j|x_k}$ and $G^{\mathbf{f}}_{jk}=\Braket{f_j|f_k}$
are Gram matrices calculated from
$\mathbf{x}^{(l)}$ and $\mathbf{f}^{(l)}$
by averaging over the sample (\ref{xfmap}).
The inverse of the Gram matrix appears in (\ref{psiXFGflocalizedAppendix}).
Whereas (\ref{psiXFGflocalizedAppendix}) may resemble the normalized components of (\ref{xfmap}),
it is used in a completely different way -- as wavefunctions, not as values.
For example, adding arbitrary $\pm$ signs in (\ref{psiXFGflocalizedAppendix}) leaves the result intact.
The idea is then to construct a quantum channel (\ref{KrausOperator}) (or (\ref{operatorTransformU}) when performing a partially unitary mapping) that maps (\ref{psiXFGflocalizedAppendix}) with maximal possible fidelity.
This approach is completely invariant with respect to an arbitrary non-degenerate linear transform of input $\mathbf{x}$
and output $\mathbf{f}$ attributes. See Appendix A of \cite{malyshkin2019radonnikodym}
and the corresponding software, where it is demonstrated that any regularization of the input and output data produces exactly the same result.

\item
Neural networks, unitary learning in the form of \cite{arjovsky2016unitary},
and most other ML knowledge representations typically work with the values of the observed data.
Instead of directly ``interpolating'' the data, we approach this in the style of
the Radon-Nikodym derivative from Ref. \cite{malyshkin2019radonnikodym},
where a state in the form of a wavefunction, such as (\ref{psiXFGflocalizedAppendix}),
is obtained first, and then the observable in this state is computed as averaged with $|\psi|^2$.
This approach separates probabilities from values: the situation is analogous to quantum mechanics,
where a state that is ``several orders off'' essentially does not affect the result if its probability is near zero.
When the states are density matrices, a highly developed theory of quantum channels can be applied to study data transformations.

\end{itemize}

The disadvantages of the approach are mostly technical.
While the developed algebraic approach is applicable to partially unitary transformations (\ref{operatorTransformU}),
its application to a general quantum channel (\ref{KrausOperator}) is still a work in progress.
There are two problems. The first is the problem of optimal parametrization of a quantum channel.
Among the most straightforward options are Kraus operators, which lead to a highly degenerate optimization problem,
and the Choi matrix, which substantially increases the problem dimension.
In Ref. \cite{belov2026semidefinitePRE}, we showed that Choi matrices reconstructed
from classical data typically have a rank of only about 1\% of the maximal possible rank.
These difficulties are primarily technical in nature and can likely be mitigated through further algorithmic improvements.

The second challenge is the requirement that the fidelity be of the Rayleigh-quotient type,
that is, expressible as a ratio of quadratic forms in the mapping operators.
This is a fundamental requirement for the applicability of the algebraic approach developed in this work.
In the general case, constructing such a fidelity is a nontrivial problem, as discussed in Section \ref{FidelityFormulation}.
Fortunately, the situation is considerably simplified by the fact that a common setting
in data analysis involves mappings from pure states of dimension $n$ to pure states of dimension $D$ (\ref{psiXFGflocalizedAppendix}).
An alternative to the algebraic approach, applicable to arbitrary fidelity functions,
is to employ gradient-based optimization methods in the spirit of neural-network training,
similar to the approaches of \cite{arjovsky2016unitary,wen2013feasible} for unitary learning.
However, such methods do not explicitly exploit the algebraic structure of the problem --
the very feature that enables both the development of more efficient numerical algorithms and a deeper understanding of the resulting solutions.
The algebraic perspective therefore offers a fundamentally different viewpoint on the use of quantum channels as a classical computational model.

In this appendix, we propose uni-directional processing, a memoryless computation.
If we introduce internal feedback loops, we allow the system to maintain a state or memory.
In Appendix C of Ref. \cite{belov2024quantumPRE}, we explored feedback examples of the mapping operator's outputs.
The current work extends that idea by assuming the internal state should take the form of a density matrix.
Practically, in a circuit like (\ref{qcExampleQCClassic}), one could introduce a feedback loop to create a system with memory.
In the density matrix network, the unit that can be passed forward or backward is a density matrix.
The topic of density matrix networks with memory will be the subject of future research.

\section{\label{QuadraticFidelity}Quadratic Form Representation of Square Root Uhlmann Fidelity for an Arbitrary Quantum Channel}
In this appendix, we show that for an arbitrary CPTP quantum channel (\ref{KrausOperator})
subject to the TP constraints (\ref{constraintKraussSpur}), the fidelity can be expressed as a quadratic form,
and the optimization problem of a sum of $M$ fidelities (\ref{fidelityStdDefinitionGeneralCase})
for individual matches can be reduced to a QCQP problem, albeit in a space of much higher dimension than the Kraus operator space.

Consider Uhlmann fidelity (\ref{fidelityDefinitionTextBook}); it corresponds to the familiar $\Braket{\phi|\varrho|\phi}$ fidelity when one of the states is pure.
The difficulty with this definition is that it does not have the interpretation of a probability,
but rather of a probability squared. If instead we consider the square root of the Uhlmann fidelity,
\begin{align}
F_{SQU}(\varrho,\sigma)&=\mathrm{Tr}\sqrt{\varrho^{1/2}\sigma\varrho^{1/2}}=
\mathrm{Tr}\left|\sqrt{\varrho}\sqrt{\sigma}\right|
\label{fidelityDefinitionTextBookSQU}
\end{align}
then the resulting sum (\ref{fidelityStdDefinitionGeneralCase}) can be interpreted as a probability.
We pursued this approach in \cite{belov2024quantumPRE}, where we considered the mapping $\sqrt{\rho}\to\sqrt{\varrho}$.
For a unitary quantum channel, the transformations of $\rho$ and $\sqrt{\rho}$ are equivalent,
since both are transformed by the same unitary operator.
Numerical simulations reported in \cite{belov2024quantumPRE}
demonstrate the exact reconstruction of a unitary mapping between density matrices
by maximizing the square root of the Holevo fidelity (\ref{fidelitySQRTvarrosigma}).
\begin{align}
F&=\mathrm{Tr}\left(\sqrt{\sigma}\sqrt{\varrho}\right)
\label{SqrtF}
\end{align}
Since no modulus (or trace norm) is taken, this quantity is generally not identical to the
Uhlmann fidelity and may provide a lower estimate of it.
The form (\ref{SqrtF}) is closely related to the
\href{https://en.wikipedia.org/wiki/Hellinger_distance}{Hellinger distance}\cite{hellinger1909neue,pollard2002user}
\begin{align}
H^2(P,Q)&=\frac{1}{2}\int\left(\sqrt{p(x)}-\sqrt{q(x)}\right)^2 dx 
\label{Hellinger} \\
&= 1-\int\sqrt{p(x)}\sqrt{q(x)} dx \nonumber
\end{align}
Representing
\begin{align}
\varrho&=PP^\dagger \label{varrhoPP}\\
\sigma&=\Sigma\Sigma^\dagger \label{sigmaSS}
\end{align}
and replacing the integral by a trace, yields
\begin{align}
H^2(\sigma,\varrho)&=\frac{1}{2}\mathrm{Tr}\left(P-\Sigma\right)\left(P^{\dagger}-\Sigma^{\dagger}\right)
\label{HellingerSigmaRho}
\end{align}
Expanding the product and choosing $P=\sqrt{\varrho}$ and $\Sigma=\sqrt{\sigma}$, obtain (\ref{SqrtF}).

The fidelity used in this appendix is precisely the square-root Uhlmann fidelity (\ref{fidelityDefinitionTextBookSQU}).
Introducing the stacked operator
\begin{align}
\Pi&=\left(
  \begin{matrix}
  B_0 P\\
  B_1 P\\
  \vdots \\
  B_{N_s-1}P
  \end{matrix}
  \right)
  \label{stackedPi}
\end{align}
which is a matrix of dimension $DN_s\times n$, we obtain
\begin{align}
\Pi\Pi^{\dagger}&=\left(
\begin{matrix}
B_0 \rho B^{\dagger}_0 & B_0 \rho B^{\dagger}_1 & \cdots \\
B_1 \rho B^{\dagger}_0 & B_1 \rho B^{\dagger}_1 & \cdots \\
\vdots & \vdots & \ddots
\end{matrix}\right)
\end{align}
The output density matrix $\varrho$ is recovered by taking the partial trace over the Kraus space:
\begin{align}
\varrho&=\mathrm{Tr}_{K} \Pi\,\Pi^{\dagger} \label{vrhpTrK}
\end{align}
Similarly, the target density matrix $\sigma$, corresponding to the desired state $\varrho^{(l)}$ in (\ref{mixedRhoMap}),
can be represented in an analogous form. Specifically,
\begin{align}
\sigma&=\mathrm{Tr}_{K} \Xi\,\Xi^{\dagger} \label{sigmaXiTrK}
\end{align}
where $\Xi$ depends only on $\varrho^{(l)}$ and on the chosen purification, and is therefore independent of the Kraus operators $B_s$.
To compare $\varrho$ (\ref{vrhpTrK})  with $\sigma$ (\ref{sigmaSS}),
we employ the Uhlmann fidelity.
Let $\Psi_{\varrho}$ and $\Psi_{\sigma}$ be purifications of $\varrho$ and $\sigma$, respectively,
in an extended Hilbert space.
Then the square-root Uhlmann fidelity (\ref{fidelityDefinitionTextBookSQU}) can be expressed as
the maximum overlap over all purifications,
$\max\limits_{\mathrm{pur}}|\Braket{\Psi_{\varrho}|\Psi_{\sigma}}|$.
Using the representations above for $\Pi$ 
and $\Xi$, we obtain
\begin{align}
F_{SQU}(\varrho(\rho^{(l)}),\varrho^{(l)})&=\max_U \mathrm{Tr}\, \Pi^{\dagger} \Xi \,U
\label{FsquAnuD}
\end{align}
where $\varrho(\rho^{(l)})$ denotes the image of $\rho^{(l)}$ under the quantum channel (\ref{KrausOperator}).
If the dimensions of the purifications differ, $\Xi$ may be padded with zeros and $U$ extended accordingly to a larger unitary matrix.
The absolute-value sign is omitted in (\ref{FsquAnuD}) because any overall phase can be absorbed into the unitary matrix $U$.
The total fidelity is then
\begin{align}
\mathcal{F}&=\sum_{l=1}^{M} \omega^{(l)}
\mathrm{Tr}\, \Pi^{\dagger} \Xi\, U^{(l)}
\label{fidelityStdDefinitionGeneralCaseUU}
\end{align}
The optimization over $U$ in (\ref{FsquAnuD}) can be incorporated into the global optimization of
$\mathcal{F}$, and the corresponding auxiliary operator is denoted by $U^{(l)}$.
Consequently, if in (\ref{fidelityStdDefinitionGeneralCaseUU}) we treat both the Kraus
operators $B_s$ and the $M$ unitary operators $U^{(l)}$ as optimization variables,
the resulting problem is exactly a QCQP.
The objective function (\ref{fidelityStdDefinitionGeneralCaseUU}) is linear in $\Pi$
(and therefore linear in the Kraus operators $B_s$) and linear in $U^{(l)}$.
Hence, it is quadratic in the optimization variables.
The constraints consist of the TP constraints (\ref{constraintKraussSpur}) together with
$M$ sets of unitarity constraints for the auxiliary operators $U^{(l)}$.
The objective function in (\ref{fidelityStdDefinitionGeneralCaseUU}) is precisely the Uhlmann fidelity,
where the matrix square root appearing in the explicit expression (\ref{fidelityDefinitionTextBookSQU})
is replaced by an optimization over auxiliary unitary operators $U^{(l)}$.

Although the formulation (\ref{fidelityStdDefinitionGeneralCaseUU}) is computationally
highly inefficient because it introduces $M$ auxiliary unitary operators $U^{(l)}$ as optimization variables,
it demonstrates that the general quantum-channel reconstruction problem can be reduced to a QCQP problem,
to which the theory developed in the present work can be applied directly.
Identifying computationally efficient formulations of this optimization
problem remains an important direction for future research.

\bibliography{LD,mla}

\end{document}